\documentstyle[11pt]{article}
\title {\bf 3N potentials in the Faddeev coordinate space
approach to Nd scattering}
\author {M.A.Braun$^{a,b}$,V.M.Suslov$^{a,b}$, I. Filikhin$^{a}$ and B.Vlahovic$^a$\\
$^a$ School of Science, NCCU, NC, USA\\
$^b$ Dep. of High Energy physics, University of S.Petersburg,\\
198504 S.Petersburg, Russia}
\def\beq{\begin{equation}}
\def\eeq{\end{equation}}

\def\vsigma{\vec{\sigma}}

\def\vx{{\bf x}}
\def\vy{{\bf y}}
\date{}
\begin{document}
\maketitle
\section{3N forces: an overview}
Most compellingly the need for the 3N forces comes from the
underbinding of light nuclei ~\cite{nogga,pieper}. Also certain
discrepancies exist in Nd scattering observables: the notorious
A$_y$-puzzle for the analyzing power.  The differential cross-sections
tensor-analyzing powers and spin-transfer coefficients are
rather well described at low energies using solely 2N forces.
However the results of these calculations start to deviate from the data as the energy 
increases ~\cite{epelbaum}.

Phenomenological 2N potential models include the CD-Bonn2000, Argonne
V-18 (AV18) and Nijmegen I and II potentials ~\cite{slaus}.
The 3N forces are represented by the Tucson-Melbourne potential and
its chirally-corrected version (TM99) ~\cite{coonhan}, Brazilian
potential ~\cite{coelho}, Urbana IX potential ~\cite{pudiner} and
and Illinois potentials ~\cite{pieper1}. All potentials contain the longest-range
 contribution due to the two-pion exchange. The TM and Brazilian potentials also 
 include the shorter -range contribution due to
$\pi\rho$ and $\rho\rho$ exchanges. In the Urbana potential this short range 
contribution is introduced phenomenologically. The Illinois models include  
parametrizations of three-pion exchange terms due to ring diagrams with one $\Delta$ 
in the intermediate state.

One should note that the phenomenological 3N forces are parametrized in the context of a 
particular 2N-force model. It is to be noted that the Urbana
3N force taken together with e AV14 2N force has been adjusted to the triton 
and $^4$He binding energies ~\cite{wiringa}and has been used in the
3N scattering calculations below the breakup threshold ~\cite{kievsky}.

A more consistent approach can be developed in the frame of chiral
perturbation theory, where both 2N and 3N forces are derived from a given
effective Lagrangian at different orders of perturbation expansion.
A detailed discussion of the structure and application of chiral nuclear forces 
may be found in the review articles ~\cite{bedaque,epelbaum1}
and references therein.

In this first attempt to include 3N forces in our calculational scheme
we shall mostly consider the simplest contribution to TM potential which
comes from the two-pion exchange.

\section{TM-$\pi\pi$ potential in the coordinate space}
The TM-$\pi\pi$ potential in the momentum space can be taken from ~\cite{coonpenna}. It reads
\[
<p_1',p_2',p_3'|W_{\pi\pi}(3)|p_1,p_2,p_3>=(2\pi)^3\frac{\delta^3(\sum p_i'-\sum p_i)}
{(q^2+\mu^2)({q'}^2+\mu^2)}
\frac{g^2}{4m^2}F^2_{\pi NN}(q)F^2_{\pi NN}(q')\]
\beq(\sigma_1q)(\sigma_2q')
\Big\{(\tau_1\tau_2)\Big[a+b(qq')\Big]-d(\tau_3\cdot\tau_1\times\tau_2)
(\sigma_3\cdot q\times q')\Big\}
\label{wmom}
\eeq
%\end{document}
Here it is assumed that nucleon 3 exchanges a pion with each of the two rest nucleons 1 and 2. $g^2 =13.4$, 
$q=p_1-p'_1$ and $q'=p'_2-p_2$. According to later studies ~\cite{coonhan}
the so-called $c$-term with $q^2+{q'}^2$ in the square brackets, which is present in
~\cite{coonpenna}, is omitted
as essentially equivalent to the first, $a$-term with a new value of $a$.
The Fourier transform gives
\[
<r_1',r_2',r_3'|W_{\pi\pi}(3)|r_1,r_2,r_3>\]\beq
=\int\prod_{i=1}^3\frac{d^3p'_i}{(2\pi)^3}\frac{d^3p_i}{(2\pi)^3}
\exp\Big(\sum_{i=1}^3i(p'_ir'_i-p_ir_i)\Big)
<p_1',p_2',p_3'|W_{\pi\pi}(3)|p_1,p_2,p_3>
\label{wcoord}
\eeq

Of the three final momenta $p'_1,p'_2$ and $p'_3$ the last can be fixed by the $\delta$-function
in (\ref{wmom}). The two others can be expressed via $q$ and $q'$ as $p'_1=p_1-q$ and $p'_2=p_2+q'$.
The exponent becomes
\[i\sum_{i=1}^3(p'_ir'_i-p_ir_i)=i\sum_{i=1}^3p_i(r'_i-r_i)-iq(r'_1-r'_3)+iq'(r'_2-r'_3)\]
%\end{document}
Integration over $p_i$, $i=1,2,3$ gives a factor
\[(2\pi)^9\delta^3(r'_1-r_1)\delta^3(r'_2-r_2)\delta^3(r'_3-r_3)\]
which makes the potential local in the coordinate space.
So finally we obtain
\[<r_1',r_2',r_3'|W_{\pi\pi}(3)|r_1,r_2,r_3>=
\delta^3(r'_1-r_1)\delta^3(r'_2-r_2)\delta^3(r'_3-r_3)\frac{g^2}{4m^2}\]\[
\int\frac{d^3q}{(2\pi)^3}\frac{d^3q'}{(2\pi)^3}e^{-iqr_{13}+iq'r_{23}}
\frac{F^2_{\pi NN}(q)}{q^2+\mu^2}\frac{F^2_{\pi NN}(q')}{{q'}^2+\mu^2}\]\beq
(\sigma_1q)(\sigma_2q')
\Big\{(\tau_1\tau_2)\Big[a+b(qq')\Big]-d(\tau_3\cdot\tau_1\times\tau_2)(\sigma_3\cdot q\times q')\Big\}
\label{wcoord1}
\eeq
where $r_{13}=r_1-r_3$ and $r_{23}=r_2-r_3$. In the following, to simplify notations, we omit
subindex 3 and write $r_{13}$ just as $r_1$ and $r_{23}$ as $r_2$ (as if the third nucleon is
located at the origin).
In terms linear or quadratic in momenta we substitute the latter by the derivatives.
Introducing also a function
\beq
Z(r)=\frac{4\pi}{\mu}\int\frac{d^3q}{(2\pi)^3}\frac {F^2_{\pi NN}(q)}{q^2+\mu^2}
\eeq
we find the expression for the (local) $W_{\pi\pi}(3)$ potential
\[
W_{\pi\pi}(3)=\frac{g^2\mu^2}{64\pi^2m^2}
(\sigma_1\nabla_1)(\sigma_2\nabla_2)\]\beq
\Big\{(\tau_1\tau_2)\Big[a+b(\nabla_1\nabla_2)\Big]-
d(\tau_3\cdot\tau_1\times\tau_2)(\sigma_3\cdot \nabla_1\times \nabla_2)\Big\}
Z(r_1)Z(r_2)
\label{wpot}
\eeq

As one knows, acting on a function of modulus,  gradients become proportional to vectors:
\[\nabla f(r)={\bf r}\frac{f'(r)}{r},\ \ \nabla_i\nabla_k f(r)=\delta_{ik}\frac{f'(r)}{r}+
r_ir_k\frac{f''(r)}{r^2}\]
Using this and introducing derivatives
\beq
Z^{(1)}=\frac{1}{r}Z'(r), \ \ Z^{(2)}=\frac{1}{r^2}Z''(r)
\eeq
we rewrite the potential in its final form
\beq
W_{\pi\pi}(3)=\frac{g^2\mu^2}{64\pi^2m^2}\Big(aW^{(a)}+bW^{(b)}+dW^{(d)}\Big)
\label{wpot1}
\eeq
where
\beq
W^{(a)}=(\tau_1\tau_2)(\sigma_1r_1)(\sigma_2r_2)Z^{(1)}(r_1)Z^{(1)}(r_2)
\label{wa}
\eeq
\[
W^{(b)}=(\tau_1\tau_2)\Big[(\sigma_1\sigma_2)Z^{(1)}(r_1)Z^{(1)}(r_2)+
(\sigma_1r_1)(\sigma_2r_1))Z^{(2)}(r_1)Z^{(1)}(r_2)\]\beq+
(\sigma_1r_2)(\sigma_2r_2))Z^{(1)}(r_1)Z^{(2)}(r_2)+
(\sigma_1r_1)(\sigma_2r_2))(r_1r_2)Z^{(2)}(r_1)Z^{(2)}(r_2)\Big]
\label{wb}
\eeq
\[
W^{(d)}=-(\tau_3\cdot \tau_1\times\tau_2)\Big[
{(\sigma_3\cdot \sigma_1\times\sigma_2)Z^{(1)}(r_1)Z^{(1)}(r_2)
+(\sigma_3\cdot \sigma_1\times r_2)(\sigma_2r_2)Z^{(1)}(r_1}Z^{(2)}(r_2)\]\beq
+(\sigma_3\cdot r_1\times \sigma_2)(\sigma_1r_1)Z^{(2)}(r_1)Z^{(1)}(r_2)
+(\sigma_3\cdot r_1\times r_2)(\sigma_1r_1)(\sigma_2r_2)Z^{(2)}(r_1)Z^{(2)}(r_2)\Big]
\label{wd}
\eeq

The latest set of parameters (TM99 ~\cite{coonhan}) is
\beq
\mu a=-1.12,\ \ \mu^3b=-2.80,\ \ \mu^3d=-0.75
\eeq
with the pion mass $\mu=138.0$ MeV and  the form-factor squared
\beq
F^2_{\pi NN}(q)=\frac{(\Lambda^2-\mu^2)^2}{(\Lambda^2+q^2)^2}
\eeq
with $\Lambda =5.8\mu$.

\section{Isospin operators in the MGL basis}
We have to calculate 4 operators
\[ (\tau_2\tau_3),\ \ (\tau_3\tau_1),\ \ (\tau_1\tau_2),\ \ (\tau_3\cdot\tau_1\times\tau_2)\]
in the MGL isospin basis formed by the three states
$\eta^{T,T_z}_{1/2,t}$ with given values of the total isospin $T$
and isospin $t$ of the pair 23:
\beq
\eta^{1/2,1/2}_{1/2,0}\equiv\eta_1,\
\eta^{1/2,1/2}_{1/2,1}\equiv\eta_2,\ \
\eta^{3/2,1/2}_{1/2,1}\equiv\eta_3.
\eeq

The simplest operator is obviously $(\tau_2\tau_3)$ which is diagonal in this basis:
\beq
(\tau_2\tau_3)_{ik}=\delta_{ik}a_i,\ \  i=1,2,3,\ \ a_1=-3,\ \ a_{2,3}=1
\label{eq2}
\eeq
To find matrix elements of $(\tau_3\tau_1)$ and $(\tau_1\tau_2)$ we recall that
permutations of nucleons 123$\to$231 and 123$\to$312 are accomplished by operators
$P^+$ and $P^-$ respectively ~\cite{laverne}
\beq
P^{I\pm}= \left ( \begin{array}{ccc}
    -1/2    & \mp\sqrt{3}/2 & 0 \\
    \pm\sqrt{3}/2 & -1/2  & 0\\
    0 & 0 & 1
    \end{array} \right )
\eeq
Note that $P^{\pm}$ are unitary:
\beq
P^+P^-=1
\eeq
If we denote $\eta_i^{(n)}$, $i=1,2,3$ the basis in which isospins of the two nucleons
different from $n$ are first summed and the isospin of nucleon $n$ is added to form the final
total isospin $T$ then our basis is $\eta_i^{(1)}$ and we have
\beq
\eta_i^{(2)}=\sum_{k}P^+_{ik}\eta^{(1)}_k,\ \ \eta_i^{(3)}=\sum_{k}P^-_{ik}\eta^{(1)}_k,
\eeq
or simply
\beq
\eta^{(2)}=P^+\eta^{(1)},\ \ \eta^{(3)}=P^-\eta^{(1)}
\label{eq3}
\eeq
Applying $P^-$ and $P^+$  to the first and second relations (\ref{eq3})
we get
\beq
\eta^{(1)}=P^-\eta^{(2)},\ \ \eta^{(1)}=P^+\eta^{(3)}
\label{eq4}
\eeq

Using these relations we find
\[
(\tau_1\tau_2)_{ik}=<\eta^{(1)}_i|(\tau_1\tau_2)|\eta_k^{(1)}>=
\sum_{i',k'}P^+_{ii'}P^+_{kk'}<\eta^{(3)}_{i'}|(\tau_1\tau_2)|\eta_{k'}^{(3)}>
\]\beq
=\sum_{i',k'}P^+_{ii'}P^+_{kk'}a_{i'}\delta_{i'k'}
=\sum_{l}P^+_{il}a_lP^-_{lk}
\eeq
where $a_i$ are defined in (\ref{eq2}).
So as a matrix
\beq
(\tau_1\tau_2)_{ik}=
\left ( \begin{array}{ccc}
    -1/2    & -\sqrt{3}/2 & 0 \\
    \sqrt{3}/2 & -1/2  & 0\\
    0 & 0 & 1
    \end{array} \right )
\left ( \begin{array}{ccc}
    -3    & 0 & 0 \\
    0 & 1  & 0\\
    0 & 0 & 1
    \end{array} \right )
 \left ( \begin{array}{ccc}
    -1/2    & \sqrt{3}/2 & 0 \\
    -\sqrt{3}/2 & -1/2  & 0\\
    0 & 0 & 1
    \end{array} \right )=
\left ( \begin{array}{ccc}
    0    & \sqrt{3} & 0 \\
    \sqrt{3} & -2  & 0\\
    0 & 0 & 1
    \end{array} \right )
\eeq
To find $\tau_3\tau_1$ we have only to substitute $P^+\leftrightarrow P^-$, which gives
\beq
(\tau_1\tau_3)_{ik}=
\left ( \begin{array}{ccc}
    0    & -\sqrt{3} & 0 \\
    -\sqrt{3} & -2  & 0\\
    0 & 0 & 1
    \end{array} \right )
\eeq

Our last operator $(\tau_3\cdot\tau_1\times\tau_2)$ is symmetric in all the three nucleons. To find its
matrix elements we use
\beq
(\tau_2\tau_1)(\tau_2\tau_3)=\tau_1\tau_3-i(\tau_3\cdot\tau_1\times\tau_2),\ \
(\tau_2\tau_3)(\tau_2\tau_1)=\tau_1\tau_3+i(\tau_3\cdot\tau_1\times\tau_2)
\eeq
wherefrom we conclude
\beq
2i(\tau_3\cdot\tau_1\times\tau_2)=(\tau_2\tau_3)(\tau_2\tau_1)-(\tau_2\tau_1)(\tau_2\tau_3)
\eeq
For the matrix elements we find from this
\beq
i(\tau_3\cdot\tau_1\times\tau_2)_{ik}=\frac{1}{2}(a_i-a_k)(\tau_2\tau_1)_{ik}
\eeq
As a matrix
\beq
i(\tau_3\cdot\tau_1\times\tau_2)=
\left ( \begin{array}{ccc}
    0    & -2\sqrt{3} & 0 \\
    2\sqrt{3} & 0  & 0\\
    0 & 0 & 0
    \end{array} \right )
\eeq

\newpage
\section{Spin-coordinate operators in the MGL basis. Generalities}
In the coordinate space our basic function is expanded in the MGL states $|\alpha>$
\beq
\sum_{\alpha}\frac{\phi_{\alpha}(x,y}{xy}|\alpha>
\eeq
The spin-coordinate part of $|\alpha>$ is labelled as
\beq
|\alpha>=|\lambda,s,l,\sigma,j>
\eeq
where we suppress the quantum numbers which are conserved: total angular momentum
$M$ its projection $M_z$ and parity $P$.
Our task is to expand in $|\alpha>$ the result of action of some operator $O({\bf x,y})$ on
our basic state
\beq
O({\bf x,y})\sum_{\alpha}\frac{\phi_{\alpha}(x,y}{xy}|\alpha>=
\sum_{\alpha}\frac{D_{\alpha}(x,y}{xy}|\alpha>
\eeq
Orthogonality of states $|\alpha>$ allows to find
\beq
D_\alpha(x,y)=\sum_\beta O_{\alpha\beta}\phi(x,y)
\eeq
where
\beq
O_{\alpha\beta}=<\alpha|O(\hat{\bf x},\hat{\bf y})|\beta>
\eeq
is the matrix element in the MGL states taken over spin and angular variables.

In the calculation of matrix elements of the 3N potential a much more
convenient is the basis $|a>$ in which individual spin projections of the nucleons
and orbital momenta of pair 23 and nucleon 1 respective this pair are given
\beq
|a>=|\lambda,\lambda_z,l,l_z,p_1,p_2,p_3>=Y_{\lambda,\lambda_z}(\hat{\bf y})
Y_{l,l_z}(\hat{\bf x})\chi^{(1)}_{p_1}\chi^{(2)}_{p_2}\chi^{(3)}_{p_3}
\eeq
Here $\chi^{(i)}_{p_i}$ is a spinor describing the $i$-th nucleon spin state
with its projection $p_i=\pm 1/2$.
The relation between the two bases is direct:
\beq
|\alpha>=\sum_a|a><a|\alpha>\equiv \sum_aF^a_\alpha|a>
\eeq
Coefficients $F^a_\alpha$ are expressed as a product of four Clebsh-Gordon
coefficients:
\beq
F^a_\alpha=C^{MM_z}_{ss_z\lambda\lambda_z}C^{ss_z}_{\frac{1}{2}p_1JJ_z}C^{JJ_z}_{\sigma\sigma_zll_z}
C^{\sigma\sigma_z}_{\frac{1}{2}p_2\frac{1}{2}p_3}
\eeq
and the sum over $a$ is in fact the sum over projections of the momenta
\beq
\sum_a=\sum_{\lambda_z,l_z,p_1,p_2,p_3}
\eeq

Calculation of matrix elements in the MGL basis then reduces to their calculation in the
factorized $a$-basis. For any operator $O$
\beq
<\alpha'|O|\alpha>=\sum_{a,a'}F^{a'}_{\alpha'}F^{a}_a<a'|O|a>
\label{atoalpha}
\eeq
(here we use that $F^a_{\alpha}$ are real).

Taking matrix elements of spin operators in the $a$-basis is trivial, since
it gives the standard Pauli matrices either in Cartesian or spherical coordinates.
Just to remind: in the Cartesian coordinates
\beq
\sigma_x=
\left ( \begin{array}{cc}
    0    & 1  \\
    1 & 0  \\
    \end{array} \right ),\ \
\sigma_y=
\left ( \begin{array}{cc}
    0    & -i  \\
    i & 0  \\
    \end{array} \right ),\ \
\sigma_z=
\left ( \begin{array}{cc}
    1    & 0  \\
    0 & -1  \\
    \end{array} \right ),\ \
\eeq
whereas in the spherical contravariant coordinates
\beq
\sigma^{+1}=-\sqrt{2}
\left ( \begin{array}{cc}
    0    & 0  \\
    1 & 0  \\
    \end{array} \right ),\ \
\sigma^0=
\left ( \begin{array}{cc}
    1    & 0  \\
    0 & -1  \\
    \end{array} \right ),\ \
\sigma^{-1}=\sqrt{2}
\left ( \begin{array}{cc}
    0    & 1  \\
    0 & 0  \\
    \end{array} \right ).
\eeq

Calculating the orbital part one encounters integrals of three spherical functions
with possible weights $\hat{\bf x}_\mu$, $\hat{\bf x}_{\mu}\hat{\bf x}_\nu$ and
$\hat{\bf x}_{\mu}\hat{\bf x}_\nu\hat{\bf x}_\epsilon$.
Correspondingly we define
\beq
B(l'l'_z\rho\tau|ll_z)\equiv
\int d\hat{\bf x}Y^*_{l'l'_z}(\hat{\bf x})Y^*_{\rho\tau}(\hat{\bf x})Y_{ll_z}(\hat{\bf x})=
\frac{1}{\sqrt{4\pi}}\frac{\prod_{l'\rho}}{\prod_l}C^{l0}_{l'0\rho 0}
C^{ll_z}_{l'l'_z\rho\tau}
\eeq
where we denote $\prod_{abc...}=[(2a+1)(2b+1)(2c+1)...]^{1/2}$.
Note that the right-hand side is real. So we  also have
\beq
B(l'l'_z\rho\tau|ll_z)=
\int d\hat{\bf x}Y_{l'l'_z}(\hat{\bf x})Y_{\rho\tau}(\hat{\bf x})Y^*_{ll_z}(\hat{\bf x})
\eeq
We use relations (Varshalovich 13.2.3.10,13.2.3.11 and 13.2.3.17)
\beq
\hat{\bf x}_\mu Y_{ll_z}(\hat{x})=\sum_{l''l''_z}\sqrt{\frac{2l+1}{2l''+1}}C^{l''l''_z}_{ll_z1\mu}
Y_{l''l''_z}(\hat{\bf x})
\label{xspher}
\eeq
and
\beq
\hat{\bf x}_\mu \hat{\bf x}_\nu Y_{ll_z}(\hat{x})=
\frac{1}{3}(-1)^\mu\delta_{\mu-\nu}Y_{ll_z}(\hat{x})+
\sum_{l''l''_z}\sqrt{\frac{2(2l+1)}{3(2l''+1)}}
C^{l''0}_{l020}C^{2\kappa}_{1\mu1\nu}C^{l''l''_z}_{ll_z2\kappa}
Y_{l''l''_z}(\hat{\bf x})
\label{xxspher}
\eeq
Multiplying (\ref{xxspher})by ${\bf x}_{\epsilon}$ and using (\ref{xspher}) we get
\[
\hat{\bf x}_\mu \hat{\bf x}_\nu \hat{\bf x}_\epsilon Y_{ll_z}(\hat{x})=
\frac{(-1)^{\mu}}{3}\delta_{\mu\nu}\sum_{l''l''_z}\sqrt{\frac{2l+1}{2l''+1}}
C^{l''l''_z}_{ll_z1\epsilon}
Y_{l''l''_z}(\hat{\bf x})\]\beq
+\sum_{l''l''_z\bar{l}''\bar{l}''_z}\sqrt{\frac{2(2l+1)}{3(2\bar{l}''+1)}}
C^{l''0}_{l020}C^{2\kappa}_{1\mu1\nu}C^{l''l''_z}_{ll_z2\kappa}
C^{\bar{l}''0}_{10l''0}C^{\bar{l}''\bar{l}''_z}_{1{\epsilon}l''l''_z}Y_{\bar{l}''\bar{l}''_z}
\eeq
Correspondingly we define
\beq
B_\mu(l'l'_z\rho\tau|ll_z)\equiv
\int d\hat{\bf x}\hat{\bf x}_\mu Y^*_{l'l'_z}(\hat{\bf x})Y^*_{\rho\tau}(\hat{\bf x})Y_{ll_z}(\hat{\bf x})
=\sum_{l''l''_z}\sqrt{\frac{2l+1}{2l''+1}}C^{l''l''_z}_{ll_z1\mu}
B(l'l'_z\rho\tau|l''l''_z)
\eeq
\[
B_{\mu\nu}(l'l'_z\rho\tau|ll_z)\equiv
\int d\hat{\bf x}\hat{\bf x}_\mu \hat{\bf x}_\nu
Y^*_{l'l'_z}(\hat{\bf x})Y^*_{\rho\tau}(\hat{\bf x})Y_{ll_z}(\hat{\bf x})\]\beq=
\frac{1}{3}(-1)^\mu\delta_{\mu-\nu}B(l'l'_z\rho\tau|ll_z)+
\sum_{l''l''_z}\sqrt{\frac{2(2l+1)}{3(2l''+1)}}
C^{l''0}_{l020}C^{2\kappa}_{1\mu1\nu}C^{l''l''_z}_{ll_z2\kappa}
B(l'l'_z\rho\tau|l''l''_z)
\eeq
and
\[
B_{\mu\nu\epsilon}(l'l'_z\rho\tau|ll_z)\equiv
\frac{(-1)^{\mu}}{3}\delta_{\mu\nu}\sum_{l''l''_z}\sqrt{\frac{2l+1}{2l''+1}}
C^{l''l''_z}_{ll_z1\epsilon}
B(l'l'_z\rho\tau|l''l''_z)\]\beq
+\sum_{l''l''_z\bar{l}''\bar{l}''_z}\sqrt{\frac{2(2l+1)}{3(2\bar{l}''+1)}}
C^{l''0}_{l020}C^{2\kappa}_{1\mu1\nu}C^{l''l''_z}_{ll_z2\kappa}
C^{\bar{l}''0}_{10l''0}C^{\bar{l}''\bar{l}''_z}_{1{\epsilon}l''l''_z}
B(l'l'_z\rho\tau|\bar{l}''\bar{l}''_z)
\eeq
Finally we recall that in our basis
\beq
{\bf r}_2-{\bf r}_3={\bf x},\ \ {\bf r}_3-{\bf r}_1=-\frac{1}{2}{\bf x}-\frac{\sqrt{3}}{2}{\bf y},
\ \ {\bf r}_1-{\bf r}_2=-\frac{1}{2}{\bf x}+\frac{\sqrt{3}}{2}{\bf y}
\eeq

\section {Operators to study}
We rewrite our 3N potential $W_{\pi\pi}(3)$ as
\[
W_{\pi\pi}(3)=\frac{g^2\mu^2}{64\pi^2m^2}\Big\{(\tau_1\tau_2)\Big[aO_2(3)+\]\beq
b\Big(O_1(3)+O_3(3)+O_4(3)+O_5(3)\Big)\Big]-d(\tau_3\cdot\tau_1\times\tau_2)\Big[
O_6(3)+O_7(3)+O_8(3)+O_9(3)\Big]\Big\}
\eeq
We have 9 different operators. For $W_{\pi\pi}(2)$ and $W_{\pi\pi}(1)$ we have to study 18
operators more $O_i(n)$ $ i=1,...9$, $n=1,2$. So in all we have 27 different operators.
Below we present a list of them in terms of our vectors ${\bf x}$ and ${\bf y}$
%%%%%%%%%%%%%%%%%%%%%%%%%%%%%%%%%%%%%%%%%%%%%%%%%%%%%%%%
\beq
O_1(3)=(\sigma_{1}\sigma_{2})Z^{(1)}(r_{13})Z^{(1)}(r_{23})
\eeq
\beq
O_1(2)=(\sigma_{1}\sigma_3)Z^{(1)}(r_{12})Z^{(1)}(r_{32)}
\eeq
\beq
O_1(1)=(\sigma_2\sigma_3)Z^{(1)}(r_{21})Z^{(1)}(r_{31})
\eeq

*******************************************************

\[
O_2(3)=(\sigma_1r_{13})(\sigma_2r_{23})Z^{(1)}(r_{13})Z^{(1)}(r_{23})\]\beq=
\Big[\frac{1}{2}(\vsigma_1\vx)(\vsigma_2\vx)+\frac{\sqrt{3}}{2}
(\vsigma_1\vy)(\vsigma_2\vx)\Big]Z^{(1)}(r_{13})Z^{(1)}(r_{23})
\eeq
\[
O_2(2)=(\sigma_3r_{32})(\sigma_1r_{12})Z^{(1)}(r_{12})Z^{(1)}(r_{32})\]\beq=
\Big[\frac{1}{2}(\vsigma_1\vx)(\vsigma_3\vx)-\frac{\sqrt{3}}{2}
(\vsigma_1\vy)(\vsigma_3\vx)\Big]Z^{(1)}(r_{12})Z^{(1)}(r_{23})
\eeq
\[
O_2(1)=(\sigma_2r_{21})(\sigma_3r_{31})Z^{(1)}(r_{21})Z^{(1)}(r_{31})\]\[=
\Big[-\frac{1}{4}(\vsigma_2\vx)(\vsigma_3\vx)-\frac{\sqrt{3}}{4}\Big(
(\vsigma_2\vx)(\vsigma_3\vy)-(\vsigma_3\vx)(\vsigma_2\vy)\Big)\]\beq
+\frac{3}{4}(\vsigma_2\vy)\vsigma_3\vy)\Big]
Z^{(1)}(r_{21})Z^{(1)}(r_{31})
\eeq

********************************************************************

\[
O_3(3)=(\sigma_1r_{13})(\sigma_2r_{13})Z^{(2)}(r_{13})Z^{(1)}(r_{23})\]\[=
\Big[\frac{1}{4}(\vsigma_1\vx)(\vsigma_2\vx)+\frac{\sqrt{3}}{4}\Big(
(\vsigma_1\vx)(\vsigma_2\vy)+(\vsigma_2\vx)(\vsigma_1\vy)\Big)\]\beq
+\frac{3}{4}(\vsigma_1\vy)\vsigma_2\vy)\Big]
Z^{(2)}(r_{13})Z^{(1)}(r_{23})
\eeq
\[
O_4(3)=(\sigma_1r_{23})(\sigma_2r_{23})Z^{(1)}(r_{13})Z^{(2)}(r_{23}))\]\beq=
(\vsigma_1\vx)(\vsigma_2\vx)Z^{(1)}(r_{13})Z^{(2)}(r_{23})
\eeq

\[
O_3(2)=(\sigma_1r_{12})(\sigma_3r_{12})Z^{(2)}(r_{12})Z^{(1)}(r_{23})\]\[=
\Big[\frac{1}{4}(\vsigma_1\vx)(\vsigma_3\vx)-\frac{\sqrt{3}}{4}\Big(
(\vsigma_1\vx)(\vsigma_3\vy)+(\vsigma_3\vx)(\vsigma_1\vy)\Big)\]\beq
+\frac{3}{4}(\vsigma_1\vy)\vsigma_3\vy)\Big]
Z^{(2)}(r_{12})Z^{(1)}(r_{23})
\eeq
\[
O_4(2)=(\sigma_1r_{32})(\sigma_3r_{32})Z^{(1)}(r_{12})Z^{(2)}(r_{23}))\]\beq=
(\vsigma_1\vx)(\vsigma_3\vx)Z^{(1)}(r_{12})Z^{(2)}(r_{23})
\eeq

\[
O_3(1)=(\sigma_2r_{21})(\sigma_3r_{21})Z^{(2)}(r_{21})Z^{(1)}(r_{31})\]\[=
\Big[\frac{1}{4}(\vsigma_2\vx)(\vsigma_3\vx)-\frac{\sqrt{3}}{4}\Big(
(\vsigma_2\vx)(\vsigma_3\vy)+(\vsigma_3\vx)(\vsigma_2\vy)\Big)\]\beq
+\frac{3}{4}(\vsigma_2\vy)\vsigma_3\vy)\Big]
Z^{(2)}(r_{12})Z^{(1)}(r_{31})
\eeq
\[
O_4(1)=(\sigma_2r_{31})(\sigma_3r_{31})Z^{(1)}(r_{21})Z^{(2)}(r_{31}))\]\[=
\Big[\frac{1}{4}(\vsigma_2\vx)(\vsigma_3\vx)+\frac{\sqrt{3}}{4}\Big(
(\vsigma_2\vx)(\vsigma_3\vy)+(\vsigma_3\vx)(\vsigma_2\vy)\Big)\]\beq
+\frac{3}{4}(\vsigma_2\vy)\vsigma_3\vy)\Big]
Z^{(1)}(r_{21})Z^{(2)}(r_{31})
\eeq

**********************************************************

\[
O_5(3)=(\sigma_1r_{13})(\sigma_2r_{23})(r_{13}r_{23})Z^{(2)}(r_{13})Z^{(2)}(r_{23})\]\beq=
\Big[\frac{1}{2}(\vsigma_1\vx)(\vsigma_2\vx)+\frac{\sqrt{3}}{2}
(\vsigma_1\vy)(\vsigma_2\vx)\Big](r_{13}r_{23})Z^{(2)}(r_{13})Z^{(2)}(r_{23})
\eeq
\[
O_5(2)=(\sigma_3r_{32})(\sigma_1r_{12})(r_{12}r_{32})Z^{(2)}(r_{12})Z^{(2)}(r_{32})\]\beq=
\Big[\frac{1}{2}(\vsigma_1\vx)(\vsigma_3\vx)-\frac{\sqrt{3}}{2}
(\vsigma_1\vy)(\vsigma_3\vx)\Big](r_{12}r_{32})Z^{(2)}(r_{12})Z^{(2)}(r_{32})
\eeq
\[
O_5(1)=(\sigma_2r_{21})(\sigma_3r_{31})(r_{21}r_{31})Z^{(2)}(r_{21})Z^{(2)}(r_{31}))\]\[=
\Big[-\frac{1}{4}(\vsigma_2\vx)(\vsigma_3\vx)-\frac{\sqrt{3}}{4}\Big(
(\vsigma_2\vx)(\vsigma_3\vy)-(\vsigma_3\vx)(\vsigma_2\vy)\Big)\]\beq
+\frac{3}{4}(\vsigma_2\vy)\vsigma_3\vy)\Big]
(r_{21}r_{31})Z^{(2)}(r_{21})Z^{(2)}(r_{31}))
\eeq

*******************************************************************************

\beq
O_6(3)=(\sigma_3\cdot\sigma_1\times\sigma_2)Z^{(1)}(r_{13})Z^{(1)}(r_{23})
\eeq
\beq
O_6(2)=(\sigma_2\cdot\sigma_3\times\sigma_1)Z^{(1)}(r_{32})Z^{(1)}(r_{12})
\eeq
\beq
O_6(1)=(\sigma_1\cdot\sigma_2\times\sigma_3)Z^{(1)}(r_{21})Z^{(1)}(r_{31})
\eeq

****************************************************************************

\[
O_7(3)=(\sigma_3\cdot \sigma_1\times r_{23})
(\sigma_2r_{23})Z^{(1)}(r_{13})Z^{(2)}(r_{23})\]\beq=
(\vsigma_3\cdot \vsigma_1\times \vx)(\vsigma_2\vx)Z^{(1)}(r_{13})Z^{(2)}(r_{23})
\eeq
\[
O_8(3)=(\sigma_3\cdot r_{13}\times\sigma_2)
(\sigma_1r_{13})Z^{(2)}(r_{13})Z^{(1)}(r_{23})\]\[=
\Big[\frac{1}{4}(\vsigma_3\cdot \vx\times\vsigma_2)(\vsigma_1\vx)+
\frac{\sqrt{3}}{4}\Big((\vsigma_3\cdot \vx\times\vsigma_2)(\vsigma_1\vy)+
(\vsigma_3\cdot \vy\times\vsigma_2)(\vsigma_1\vx)\Big)\]\beq+\frac{3}{4}
(\vsigma_3\cdot \vy\times\vsigma_2)(\vsigma_1\vy)\Big]
Z^{(2)}(r_{13})Z^{(1)}(r_{23})
\eeq
\[
O_7(2)=(\sigma_2\cdot \sigma_3\times r_{12})
(\sigma_1r_{12})Z^{(1)}(r_{23})Z^{(2)}(r_{12})\]\[=
\Big[\frac{1}{4}(\vsigma_2\cdot \vsigma_3\times\vx)(\vsigma_1\vx)-
\frac{\sqrt{3}}{4}\Big((\vsigma_2\cdot \vsigma_3\times\vx)(\vsigma_1\vy)+
(\vsigma_2\cdot \vsigma_3\times\vy)(\vsigma_1\vx)\Big)\]\beq+\frac{3}{4}
(\vsigma_3\cdot \vsigma_3\times\vy)(\vsigma_1\vy)\Big]
Z^{(1)}(r_{23})Z^{(2)}(r_{13})
\eeq
\[
O_8(2)=(\sigma_2\cdot r_{32}\times\sigma_1)
(\sigma_3r_{32})Z^{(2)}(r_{32})Z^{(1)}(r_{12})\]\beq=
(\vsigma_2\cdot \vx\times\vsigma_1)(\vsigma_3\vx)Z^{(2)}(r_{32})Z^{(1)}(r_{12}
\eeq
\[
O_7(1)=(\sigma_1\cdot \sigma_2\times r_{31})
(\sigma_3r_{31})Z^{(1)}(r_{23})Z^{(2)}(r_{31})\]\[=
\Big[\frac{1}{4}(\vsigma_1\cdot \vsigma_2\times\vx)(\vsigma_3\vx)+
\frac{\sqrt{3}}{4}\Big((\vsigma_1\cdot \vsigma_2\times\vx)(\vsigma_3\vy)+
(\vsigma_1\cdot \vsigma_2\times\vy)(\vsigma_3\vx)\Big)\]\beq+\frac{3}{4}
(\vsigma_1\cdot \vsigma_2\times\vy)(\vsigma_3\vy)\Big]
Z^{(1)}(r_{23})Z^{(2)}(r_{31})
\eeq
\[
O_8(1)=(\sigma_1\cdot r_{21}\times\sigma_3)
(\sigma_2r_{21})Z^{(2)}(r_{21})Z^{(1)}(r_{31})\]\[=
\Big[\frac{1}{4}(\vsigma_1\cdot \vx\times\vsigma_3)(\vsigma_2\vx)-
\frac{\sqrt{3}}{4}\Big((\vsigma_1\cdot \vx\times\vsigma_3)(\vsigma_2\vy)+
(\vsigma_1\cdot \vy\times\vsigma_3)(\vsigma_2\vx)\Big)\]\beq+\frac{3}{4}
(\vsigma_1\cdot \vy\times\vsigma_3)(\vsigma_2\vy)\Big]
Z^{(2)}(r_{21})Z^{(1)}(r_{31})
\eeq
********************************************************************

\[
O_9(3)=(\sigma_3\cdot r_{13}\times r_{23})(\sigma_1r_{13})
(\sigma_2r_{23})Z^{(2)}(r_{13})Z^{(2)}(r_{23})\]\beq=
(\vsigma_3\cdot\vy\times\vx)
\Big[\frac{\sqrt{3}}{4}(\vsigma_1\vx)(\vsigma_2\vx)+
\frac{3}{4}(\vsigma_1\vy)(\vsigma_2\vx)\Big]
Z^{(2)}(r_{13})Z^{(2)}(r_{23})
\eeq
\[
O_9(2)=(\sigma_2\cdot r_{32}\times r_{12})(\sigma_3r_{32})
(\sigma_1r_{12})Z^{(2)}(r_{32})Z^{(2)}(r_{12})\]\beq=
(\vsigma_2\cdot\vy\times\vx)
\Big[\frac{\sqrt{3}}{4}(\vsigma_3\vx)(\vsigma_1\vx)-
\frac{3}{4}(\vsigma_1\vy)(\vsigma_3\vx)\Big]
Z^{(2)}(r_{32})Z^{(2)}(r_{12})
\eeq
\[
O_9(1)=(\sigma_1\cdot r_{21}\times r_{31})(\sigma_2r_{21})(\sigma_3r_{31})
Z^{(2)}(r_{21})Z^{(2)}(r_{31})\]\[=
(\vsigma_1\cdot\vy\times\vx)
\Big[-\frac{\sqrt{3}}{8}(\vsigma_2\vx)(\vsigma_3\vx)+\frac{3}{8}
\Big((\vsigma_3\vx)(\vsigma_2\vy)-(\vsigma_2\vx)(\vsigma_3\vy)\Big)\]
\beq
+\frac{3\sqrt{3}}{8}
(\vsigma_2\vy)(\vsigma_3\vy)\Big]
Z^{(2)}(r_{21})Z^{(2)}(r_{31})
\eeq

%%%%%%%%%%%%%%%%%%%%%%%%%%%%%%%%%%%%%%%%%%%%%%%%%%%%%%%

Our idea is first to construct matrix elements of these operators in the $a$-basis, which is almost trivial
and only needs a good bookkeeping. In principle the desired matrix elements in the MGL basis can then be
obtained by Eq. (\ref{atoalpha}). Written in terms of Clebsh-Gordon coefficients this includes
multiple (though finite) summations over momentum projections. It is to be seen whether some
of these summations can be done analytically (as for Coulomb matrix elements). This  problem
is to be treated at the next step.

\section{Spin-coordinate operators in the $a$-basis}
\subsection{Operators $O_1$}
We illustrate our technique  in the simplest case of operator $O_1(3)$
As we mentioned, the spin part is trivial
\beq
<p'_1p'_2p'_3|(\sigma_1\sigma_2)|p_1p_2p_3>=
\vec{\sigma}_{p'_1p_1}\vec{\sigma}_{p'_2p_2}\delta_{p'_3p_3}
\eeq
Now the spatial part
\[
Z^{(1)}(|\frac{1}{2}{\bf x}+\frac{\sqrt{3}}{2}{\bf y}|)Z^{(1)}(x)\]
As with the Coulomb matrix elements we expand the first factor in Legendre polynomials
in $\cos \theta$ where $\theta$ is the angle between $\bf x$ and $\bf y$ and then convert this expansion
into the one in bipolar harmonics. Having in mind permutations $2\leftrightarrow 3$ we expand
\beq
Z^{(1)}(|\frac{1}{2}{\bf x}\pm\frac{\sqrt{3}}{2}{\bf y}|)=
\sum_\rho(2l+1)P_\rho(\cos\theta)(\mp 1)^\rho Z^{(1)}_\rho(x,y)=
4\pi\sum_{\rho,\tau}Y^*(\hat{\bf x})_{\rho\tau}Y(\hat{\bf y})_{\rho\tau}(\mp 1)^\rho Z^{(1)}_\rho(x,y)
\eeq
Taking matrix elements will give two $B$ coefficients
\beq
<\lambda'\lambda'_z l'l'_z|Z^{(1)}(|\frac{1}{2}{\bf x}\pm\frac{\sqrt{3}}{2}{\bf y}|)
|\lambda\lambda_zll_z>=
4\pi\sum_{\rho\tau}(\mp 1)^\rho Z^{(1)}_\rho(x,y)B(l'l'_z\rho\tau|ll_z)
B(\lambda\lambda_z\rho\tau|\lambda'\lambda'_z)
\eeq
So finally we get
\beq
<a'|O_1(3)|a>=4\pi\vec{\sigma}_{p'_1p_1}\vec{\sigma}_{p'_2p_2}\delta_{p'_3p_3}
Z^{(1)}(x)\sum_{\rho\tau}(-1)^\rho
Z^{(1)}_\rho(x,y)B(l'l'_z\rho\tau|ll_z)
B(\lambda\lambda_z\rho\tau|\lambda'\lambda'_z)
\eeq

This explicit expression is difficult to read and write. So in the following we use
a convenient short-hand matrix notation. We shall not write out the components
of the matrix element of Pauli matrices but leave them as they stand:
\[(\vec{\sigma_1}\vec{\sigma}_2)\]
As to the spatial part, we introduce  matrices
\beq
<l'l'_z|B^{\tau}(\rho)|l,l_z>\equiv B(l'l'_z\rho\tau|ll_z),\ \
<\lambda'\lambda'_z|D^{\tau}(\rho)|l,l_z>\equiv B(\lambda\lambda_z\rho\tau|\lambda'\lambda'_z)
\eeq
and their scalar product
\beq
B(\rho)D(\rho)\equiv\sum_{\tau}B^{\tau}(\rho)D^{\tau}(\rho)\equiv (B D)_{\rho}
\eeq
In these notations
\beq
<a'|O_1(3)|a>=<a'|4\pi(\vec{\sigma_1}\vec{\sigma}_2)Z^{(1)}(x)\sum_{\rho}
(-1)^\rho Z^{(1)}_\rho(x,y)(BD)_{\rho}|a>
\eeq
or simply, omitting the states $a$ and $a'$
\beq
O_1(3)=4\pi(\vec{\sigma_1}\vec{\sigma}_2)Z^{(1)}(x)\sum_{\rho}
(-1)^\rho Z^{(1)}_\rho(x,y)(BD)_{\rho}
\eeq
Permutation $2\leftrightarrow 3$ will obviously give
\beq
O_1(2)=
4\pi(\vec{\sigma_1}\vec{\sigma}_3)Z^{(1)}(x)\sum_{\rho}
 Z^{(1)}_\rho(x,y)(BD)_{\rho}
\eeq

In $O_1(1)$ the spatial part is
\[
Z^{(1)}(|\frac{1}{2}{\bf x}+\frac{\sqrt{3}}{2}{\bf y}|)
Z^{(1)}(|\frac{1}{2}{\bf x}-\frac{\sqrt{3}}{2}{\bf y}|)\]
We  expand
it as a whole in Legendre polynomials
\[
Z^{(1)}(|\frac{1}{2}{\bf x}+\frac{\sqrt{3}}{2}{\bf y}|)
Z^{(1)}(|\frac{1}{2}{\bf x}-\frac{\sqrt{3}}{2}{\bf y}|)
\]\beq=
\sum_\rho(2l+1)P_\rho(\cos\theta) Z^{(11)}_\rho(x,y)=
4\pi\sum_{\rho,\tau}Y^*(\hat{\bf x})_{\rho\tau}Y(\hat{\bf y})_{\rho\tau}Z^{(11)}_\rho(x,y)
\label{prodz}
\eeq
Obviously only even value of $\rho$ contribute.
After that we find
\beq
O_1(1)=4\pi(\vec{\sigma}_2\vec{\sigma}_3)
\sum_{\rho}Z^{(11)}_\rho(x,y)(BD)_\rho
\eeq

\subsection{Operators $O_2-O_5$}
Operators $O-2- O_5$ can be
written in a straightforward  manner using our list of operators in
terms of vectors ${\bf x}$ and ${\bf y}$ and attaching subindexes $\mu$ and $\nu$
to matrices $B$ or $D$.
\beq
O_2(3)=4\pi\sum_{\mu\nu}\sigma_1^{\mu}\sigma_2^{\nu}xZ^{(1)}(x)\sum_{\rho}
Z^{(1)}_{\rho}(x,y)(-1)^{\rho}\Big[\frac{1}{2}x(B_{\mu\nu}D)_{\rho}+\frac{\sqrt{3}}{2}y
(B_{\nu}D_{\mu})_{\rho}\Big]
\eeq

\beq
O_2(2)=4\pi\sum_{\mu\nu}\sigma_1^{\mu}\sigma_3^{\nu}xZ^{(1)}(x)\sum_{\rho}
Z^{(1)}_{\rho}(x,y)\Big[\frac{1}{2}x(B_{\mu\nu}D)_{\rho}-\frac{\sqrt{3}}{2}y
(B_{\nu}D_{\mu})_{\rho}\Big]
\eeq

\beq
O_2(1)=4\pi\sum_{\mu\nu}\sigma_2^{\mu}\sigma_3^{\nu}\sum_{\rho}
Z^{(11)}_{\rho}(x,y)\Big[-\frac{1}{4}x^2(B_{\mu\nu}D)_{\rho}-\frac{\sqrt{3}}{4}xy
\Big(B_{\mu}D_{\nu})_{\rho}-B_{\nu}D_{\mu})_{\rho}\Big)+\frac{3}{4}
y^2(BD_{\mu\nu})_{\rho}\Big]
\eeq

\[
O_3(3)=4\pi\sum_{\mu\nu}\sigma_1^{\mu}\sigma_2^{\nu}Z^{(1)}(x)\sum_{\rho}
Z^{(2)}_{\rho}(x,y)(-1)^{\rho}\Big[\frac{1}{4}x^2(B_{\mu\nu}D)_{\rho}
\]\beq+\frac{\sqrt{3}}{4}xy
\Big(B_{\mu}D_{\nu})_{\rho}+B_{\nu}D_{\mu})_{\rho}\Big)+\frac{3}{4}
y^2(BD_{\mu\nu})_{\rho}\Big]
\eeq

\[
O_3(2)=4\pi\sum_{\mu\nu}\sigma_1^{\mu}\sigma_3^{\nu}Z^{(1)}(x)\sum_{\rho}
Z^{(2)}_{\rho}(x,y)\Big[\frac{1}{4}x^2(B_{\mu\nu}D)_{\rho}\]\beq-
\frac{\sqrt{3}}{4}xy
\Big(B_{\mu}D_{\nu})_{\rho}+B_{\nu}D_{\mu})_{\rho}\Big)+\frac{3}{4}
y^2(BD_{\mu\nu})_{\rho}\Big]
\eeq

\[
O_3(1)=4\pi\sum_{\mu\nu}\sigma_2^{\mu}\sigma_3^{\nu}\sum_{\rho}
Z^{(21)}_{\rho}(x,y)\Big[\frac{1}{4}x^2(B_{\mu\nu}D)_{\rho}\]\beq-\frac{\sqrt{3}}{4}xy
\Big(B_{\mu}D_{\nu})_{\rho}+B_{\nu}D_{\mu})_{\rho}\Big)+\frac{3}{4}y^2
(BD_{\mu\nu})_{\rho}\Big]
\eeq

\beq
O_4(3)=
4\pi\sum_{\mu\nu}\sigma_1^{\mu}\sigma_2^{\nu}x^2Z^{(2)}(x)\sum_{\rho}
Z^{(1)}_{\rho}(x,y)(-1)^{\rho}(B_{\mu\nu}D)_{\rho}
\eeq

\beq
O_4(2)=
4\pi\sum_{\mu\nu}\sigma_1^{\mu}\sigma_3^{\nu}x^2Z^{(2)}(x)\sum_{\rho}
Z^{(1)}_{\rho}(x,y)(B_{\mu\nu}D)_{\rho}
\eeq

\beq
O_4(1)=4\pi\sum_{\mu\nu}\sigma_2^{\mu}\sigma_3^{\nu}\sum_{\rho}
Z^{(12)}_{\rho}(x,y)\Big[\frac{1}{4}x^2(B_{\mu\nu}D)_{\rho}+\frac{\sqrt{3}}{4}xy
\Big(B_{\mu}D_{\nu})_{\rho}+B_{\nu}D_{\mu})_{\rho}\Big)+\frac{3}{4}y^2
(BD_{\mu\nu})_{\rho}\Big]
\eeq

\beq
O_5(3)=4\pi\sum_{\mu\nu}\sigma_1^{\mu}\sigma_2^{\nu}xZ^{(2)}(x)\sum_{\rho}
U^{(2)}_{\rho}(x,y)(-1)^{\rho}\Big[\frac{1}{2}x(B_{\mu\nu}D)_{\rho}+\frac{\sqrt{3}}{2}y
(B_{\nu}D_{\mu})_{\rho}\Big]
\eeq

\beq
O_5(2)=4\pi\sum_{\mu\nu}\sigma_1^{\mu}\sigma_2^{\nu}xZ^{(2)}(x)\sum_{\rho}
U^{(2)}_{\rho}(x,y)\Big[\frac{1}{2}x(B_{\mu\nu}D)_{\rho}-\frac{\sqrt{3}}{2}y
(B_{\nu}D_{\mu})_{\rho}\Big]
\eeq

\beq
O_5(1)=4\pi\sum_{\mu\nu}\sigma_2^{\mu}\sigma_3^{\nu}{x}\sum_{\rho}
U^{(22)}_{\rho}(x,y)\Big[-\frac{1}{4}x^2(B_{\mu\nu}D)_{\rho}-\frac{\sqrt{3}}{4}xy
\Big(B_{\mu}D_{\nu})_{\rho}-B_{\nu}D_{\mu})_{\rho}\Big)+\frac{3}{4}y^2
(BD_{\mu\nu})_{\rho}\Big]
\eeq

In operators $O_5$ we introduce functions of the angle $({\bf x}{\bf y})$
\beq
U^{(1)}=({\bf r}_{12}{\bf r}_{32})Z^{(2)}(r_{12}),\ \
U^{(22)}=({\bf r}_{21}{\bf r}_{31})Z^{(2)}(r_{21)}Z^{(2)}(r_{31})
\eeq
Also $Z^{(12)}_\rho$ and $Z^{(21)}_{\rho}$ are the partial of functions
$Z^{(1)}(r_{12})Z^{(2)}(r_{13})$ and $Z^{(2)}(r_{12})Z^{(1)}(r_{13})$ respectively.
They differ by factor $(-1)^{\rho}$
\section{Operators $O_6-O_9$}
Operators $O_6$ are simple

\beq
O_6(3)=4\pi(\vsigma_3\cdot\vsigma_1\times\vsigma_2)
Z^{(1)}(x)\sum_{\rho}
(-1)^\rho Z^{(1)}_\rho(x,y)(BD)_{\rho}
\eeq

\beq
O_6(2)=
4\pi (\vsigma_2\cdot\vsigma_3\times\vsigma_1)Z^{(1)}(x)\sum_{\rho}
 Z^{(1)}_\rho(x,y)(BD)_{\rho}
\eeq

\beq
O_6(1)=
4\pi (\vsigma_1\cdot\vsigma_2\times\vsigma_3)
\sum_{\rho} Z^{(11)}_\rho(x,y)(BD)_{\rho}
\eeq

To write out other operators we use the form of the vector product in
spherical coordinates:
\beq
[a\times b]^{\epsilon}=i\sqrt{2}\sum_{\mu\nu}C^{1\epsilon}_{1\mu 1\nu}a^{\mu}b^{\nu}
\eeq
The we easily find
\beq
O_7(3)=4\pi i\sqrt{2}\sum_{\mu\nu\epsilon\phi}C^{1\epsilon}_{1\mu 1\nu}
\sigma_3^{\epsilon}\sigma_1^{\mu}\sigma_2^{\phi}x^2Z^{(2)}(x)\sum_{\rho}
(-1)^\rho Z^{(1)}_\rho(x,y)(B_{\nu\phi}D)_{\rho}
\eeq

\[
O_7(2)=4\pi i\sqrt{2}\sum_{\mu\nu\epsilon\phi}C^{1\epsilon}_{1\mu 1\nu}
\sigma_2^{\epsilon}\sigma_3^{\mu}\sigma_1^{\phi}Z^{(1)}(x)\sum_{\rho}
 Z^{(2)}_\rho(x,y)\Big[\frac{1}{4}x^2(B_{\nu\phi}D)_{\rho}\]\beq
 -\frac{\sqrt{3}}{4}xy\Big((B_{\nu}D_{\phi})_{\rho}+(B_{\phi}D_{\nu})_{\rho}\Big)
 +\frac{3}{4}y^2(BD_{\nu\phi})_{\rho}\Big]
\eeq

\[
O_7(1)=4\pi i\sqrt{2}\sum_{\mu\nu\epsilon\phi}C^{1\epsilon}_{1\mu 1\nu}
\sigma_1^{\epsilon}\sigma_2^{\mu}\sigma_3^{\phi}\sum_{\rho}
 Z^{(12)}_\rho(x,y)\Big[\frac{1}{4}x^2(B_{\nu\phi}D)_{\rho}\]\beq
 +\frac{\sqrt{3}}{4}xy\Big((B_{\nu}D_{\phi})_{\rho}+(B_{\phi}D_{\nu})_{\rho}\Big)
 +\frac{3}{4}y^2(BD_{\nu\phi})_{\rho}\Big]
\eeq

\[
O_8(3)=-4\pi i\sqrt{2}\sum_{\mu\nu\epsilon\phi}C^{1\epsilon}_{1\mu 1\nu}
\sigma_3^{\epsilon}\sigma_2^{\mu}\sigma_1^{\phi}Z^{(1)}(x)\sum_{\rho}
 (-1)^{\rho}Z^{(2)}_\rho(x,y)\Big[\frac{1}{4}x^2(B_{\nu\phi}D)_{\rho}\]\beq
 +\frac{\sqrt{3}}{4}xy\Big((B_{\nu}D_{\phi})_{\rho}+(B_{\phi}D_{\nu})_{\rho}\Big)
 +\frac{3}{4}y^2(BD_{\nu\phi})_{\rho}\Big]
\eeq

\beq
O_8(2)=-4\pi i\sqrt{2}\sum_{\mu\nu\epsilon\phi}C^{1\epsilon}_{1\mu 1\nu}
\sigma_2^{\epsilon}\sigma_3^{\mu}\sigma_1^{\phi}x^2Z^{(2)}(x)\sum_{\rho}
Z^{(1)}_\rho(x,y)(B_{\nu\phi}D)_{\rho}
\eeq

\[
O_8(1)=-4\pi i\sqrt{2}\sum_{\mu\nu\epsilon\phi}C^{1\epsilon}_{1\mu 1\nu}
\sigma_1^{\epsilon}\sigma_3^{\mu}\sigma_2^{\phi}\sum_{\rho}
Z^{(21)}_\rho(x,y)\Big[\frac{1}{4}x^2(B_{\nu\phi}D)_{\rho}\]\beq
 -\frac{\sqrt{3}}{4}xy\Big((B_{\nu}D_{\phi})_{\rho}+(B_{\phi}D_{\nu})_{\rho}\Big)
 +\frac{3}{4}y^2(BD_{\nu\phi})_{\rho}\Big]
\eeq

\[
O_9(3)=4\pi i\sqrt{2}\sum_{\mu\nu\epsilon\phi\chi}C^{1\epsilon}_{1\mu 1\nu}
\sigma_3^{\epsilon}\sigma_1^{\phi}\sigma_2^{\chi}x^2Z^{(2)}(x)\]\beq
\sum_{\rho}(-1)^{\rho}Z^{(2)}_{\rho}(x,y)\Big(\frac{\sqrt{3}}{4}x(B_{\nu\phi\chi}D_{\mu})_{\rho}
+\frac{3}{4}y(B_{\nu\chi}D_{\mu\phi})_{\rho}\Big)
\eeq

\beq
O_9(2)=4\pi i\sqrt{2}\sum_{\mu\nu\epsilon\phi\chi}C^{1\epsilon}_{1\mu 1\nu}
\sigma_2^{\epsilon}\sigma_3^{\phi}\sigma_1^{\chi}x^2Z^{(2)}(x)
\sum_{\rho}Z^{(2)}_{\rho}(x,y)\Big(\frac{\sqrt{3}}{4}x(B_{\nu\phi\chi}D_{\mu})_{\rho}
-\frac{3}{4}y(B_{\nu\chi}D_{\mu\phi})_{\rho}\Big)
\eeq

\[
O_9(1)=4\pi i\sqrt{2}\sum_{\mu\nu\epsilon\phi\chi}C^{1\epsilon}_{1\mu 1\nu}
\sigma_1^{\epsilon}\sigma_2^{\phi}\sigma_3^{\chi}xy\]\[
\sum_{\rho}Z^{(22)}_{\rho}(x,y)\Big[-\frac{\sqrt{3}}{8}x^2(B_{\nu\phi\chi}D_{\mu})_{\rho}
\]\beq+\frac{3}{8}xy\Big(B_{\nu\chi}D_{\mu\phi})_{\rho}-B_{\nu\phi}D_{\mu\chi})_{\rho}\Big)
+\frac{3\sqrt{3}}{8}y^2(B_{\nu}D_{\mu\phi\chi})_{\rho}\Big]
\eeq

\section{Functions to be expanded in partial waves}
Our basic functions can be taken from ~\cite{coonpenna}
Take the form-factor as
\beq
F_{\pi\pi}(q)=\frac{\Lambda^2-\mu^2}{\Lambda^2-q^2}
\eeq
Define
\beq
G(r)=\frac{e^{-r}}{r}\Big(1+\frac{1}{r}\Big),\ \
F(r)=\frac{e^{-r}}{r}\Big(1+\frac{3}{r}+\frac{3}{r^2}\Big)
\eeq
Then
\beq
Z^{(1)}(r)=-\mu\Big[G(\mu r)-\frac{\Lambda^2}{\mu^2}G(\Lambda r)
-\frac{1}{2}\Big(\frac{\Lambda^2}{\mu^2}-1\Big)e^{-\Lambda r}\Big]
\eeq
and
\beq
Z^{(2)}(r)=\frac{1}{r}Z^{(1)}(r)+X_2(r)
\eeq
where
\beq
X_2(r)=\mu^2\Big[F(\mu r)-\frac{\Lambda^3}{\mu^3}F(\Lambda r)
-\frac{1}{2}\frac{\Lambda^2 r}{\mu}\Big(\frac{\Lambda^2}{\mu^2}-1\Big)
G(\Lambda r)\Big]
\eeq

The angular dependence comes from the expressions for $r_{12}$ and $r_{13}$:
\beq
r_{12}=\sqrt{\frac{1}{4}x^2+\frac{3}{4}y^2-\frac{\sqrt{3}}{2}xy\cos\theta},\ \
r_{13}=\sqrt{\frac{1}{4}x^2+\frac{3}{4}y^2+\frac{\sqrt{3}}{2}xy\cos\theta}
\eeq
For any function $F$ partial waves of $F(r_{12})$ and $F(r_{13})$ will differ
by a factor $(-1)^{\rho}$. We standardly define partial wave for
$F(r_{12})$:
\beq
F_{\rho}(x,y)=\frac{1}{2}\int_{-1}^{+1}d\cos\theta P_{\rho}(\cos\theta)F(r_{12})
\eeq
For $F(r_{13})$ partial waves will then be $(-1)^{\rho} F_{\rho}(x,y)$
The product of two identical functions $F(r_{12})F(r_{13})$ is even in $\cos\theta$
and its partial wave expansion will contain only even values of $\rho$.
When a product of different functions is partial wave expanded, then partial
waves $Z^{(ik)}$ correspond to the product $Z^{(i)}(r_{12})Z^{(k)}(r_{13})$

\section{Dynamical and geometrical factors}
Our next step is present our matrix elements in terms of dynamical
factors, depending on $x$ and $y$ and geometrical ones, which depend only on the
labels of the states between which the matrix elements are taken. Their role
is different in the calculational procedure. Geometrical factors can be
calculated once and for all before the actual computation, which is based
on scanning the grid in $x,y$ plane.

To simplify notations, from now on we denote
\beq
O_i(k)=O_{ik},\ \  i=1,...9,\ \ k=1,2,3
\eeq
We present
\beq
O_{ik}=\sum_{l=1}^{l_{ik}}\sum_{\rho}O_{ik\rho}^{(l)}(x,y)q_{ik\rho}^{(l)}
\label{dygeo}
\eeq
where $O_{ik\rho}^{(l)}(x,y)$ are dynamical factors and matrix elements
of $q_{ik\rho}^{(l)}$ give geometrical factors. The number of terms $\l_{ik}$
maybe 1, 2 or 3 for different $ik$.

Expansion (\ref{dygeo}) can be made in a straightforward manner on the inspection
of expressions for the operators in Section 6 .
\beq
O_{13\rho}=Z^{(1)}(x)Z^{(1)}_{\rho}(x,y)(-1)^{\rho},\ \
q_{13\rho}=4\pi(\vsigma_1\vsigma_2)(BD)_{\rho}
\eeq

\beq
O_{12\rho}=Z^{(1)}(x)Z^{(1)}_{\rho}(x,y),\ \
q_{12\rho}=4\pi(\vsigma_1\vsigma_3)(BD)_{\rho}
\eeq

\beq
O_{11\rho}=z^{(11)}_{\rho}(x,y),\ \
q_{11\rho}=4\pi(\vsigma_2\vsigma_3)(BD)_{\rho}
\eeq

\beq
O_{23\rho}^{(1)}=\frac{1}{2}
x^2Z^{(1)}(x)Z^{(1)}_{\rho}(x,y)(-1)^{\rho},\ \
q_{23\rho}^{(1)}=4\pi\sum_{\mu\nu}\sigma_1^{\mu}\sigma_2^{\nu}(B_{\mu\nu}D)_{\rho}
\eeq
\beq
O_{23\rho}^{(2)}=\frac{\sqrt{3}}{2}
xyZ^{(1)}(x)Z^{(1)}_{\rho}(x,y)(-1)^{\rho},\ \
q_{23\rho}^{(2)}=4\pi\sum_{\mu\nu}\sigma_1^{\mu}\sigma_2^{\nu}(B_{\nu}D_{\mu})_{\rho}
\eeq

\beq
O_{22\rho}^{(1)}=(-1)^{\rho}O_{23\rho}^{(1)}
\ \
q_{22\rho}^{(1)}=4\pi\sum_{\mu\nu}\sigma_1^{\mu}\sigma_3^{\nu}(B_{\mu\nu}D)_{\rho}
\eeq
\beq
O_{22\rho}^{(2)}=-(-1)^{\rho}O_{23\rho}^{(1)},\ \
q_{22\rho}^{(2)}=4\pi\sum_{\mu\nu}\sigma_1^{\mu}\sigma_3^{\nu}(B_{\nu}D_{\mu})_{\rho}
\eeq

\beq
O_{21\rho}^{(1)}=-\frac{1}{4}x^2Z^{(11)}_{\rho}(x,y),\ \
q_{21\rho}=4\pi\sum_{\mu\nu}\sigma_2^{\mu}\sigma_3^{\nu}(B_{\mu\nu}D)_{\rho}
\eeq
\beq
O_{21\rho}^{(2)}=-\frac{\sqrt{3}}{4}xyZ^{(11)}_{\rho}(x,y),\ \
q_{21\rho}^{(2)}=4\pi\sum_{\mu\nu}\sigma_2^{\mu}\sigma_3^{\nu}(B_{\mu}D_{\mu}-B_{\nu}D_{\mu})_{\rho}
\eeq
\beq
O_{21\rho}^{(3)}=\frac{3}{4}y^2Z^{(11)}_{\rho}(x,y),\ \
q_{21\rho}^{(3)}=4\pi\sum_{\mu\nu}\sigma_2^{\mu}\sigma_3^{\nu}(BD_{\mu\nu})_{\rho}
\eeq

\beq
O_{33\rho}^{(1)}=\frac{1}{4}x^2Z^{(1)}(x)Z^{(2)}_{\rho}(x,y)(-1)^{\rho},\ \
q_{33\rho}^{(1)}=q_{23\rho}^{(1)}
\eeq
\beq
O_{33\rho}^{(2)}=\frac{\sqrt{3}}{4}xyZ^{(1)}(x)Z^{(2)}_{\rho}(x,y)(-1)^{\rho},\ \
q_{33\rho}^{(2)}=4\pi\sum_{\mu\nu}\sigma_1^{\mu}\sigma_2^{\nu}(B_{\mu}D_{\mu}+B_{\nu}D_{\mu})_{\rho}
\eeq
\beq
O_{33\rho}^{(3)}=\frac{3}{4}y^2Z{(1)}(x)Z^{(2)}_{\rho}(x,y)(-1)^{\rho},\ \
q_{33\rho}^{(3)}=4\pi\sum_{\mu\nu}\sigma_1^{\mu}\sigma_2^{\nu}(BD_{\mu\nu})_{\rho}
\eeq

\beq
O_{32\rho}^{(1)}=\frac{1}{4}x^2Z^{(1)}(x)Z^{(2)}_{\rho}(x,y),\ \
q_{32\rho}^{(1)}=q_{22\rho}^{(1)}
\eeq
\beq
O_{32\rho}^{(2)}=-\frac{\sqrt{3}}{4}xyZ^{(1)}(x)Z^{(2)}_{\rho}(x,y),\ \
q_{32\rho}^{(2)}=4\pi\sum_{\mu\nu}\sigma_1^{\mu}\sigma_3^{\nu}(B_{\mu}D_{\mu}+B_{\nu}D_{\mu})_{\rho}
\eeq
\beq
O_{32\rho}^{(3)}=\frac{3}{4}y^2Z{(1)}(x)Z^{(2)}_{\rho}(x,y),\ \
q_{32\rho}^{(3)}=4\pi\sum_{\mu\nu}\sigma_1^{\mu}\sigma_3^{\nu}(BD_{\mu\nu})_{\rho}
\eeq

\beq
O_{31\rho}^{(1)}=\frac{1}{4}x^2Z^{(21)}_{\rho}(x,y),\ \
q_{31\rho}^{(1)}=q_{21\rho}^{(1)}
\eeq
\beq
O_{31\rho}^{(2)}=-\frac{\sqrt{3}}{4}xyZ^{(21)}_{\rho}(x,y),\ \
q_{31\rho}^{(2)}=4\pi\sum_{\mu\nu}\sigma_2^{\mu}\sigma_3^{\nu}(B_{\mu}D_{\mu}+B_{\nu}D_{\mu})_{\rho}
\eeq
\beq
O_{31\rho}^{(3)}=\frac{3}{4}y^2Z^{(21)}_{\rho}(x,y),\ \
q_{31\rho}^{(3)}=q_{21\rho}^{(3)}
\eeq

\beq
O_{43\rho}=x^2Z^{(2)}(x)Z^{(1)}_{\rho}(x,y)(-1)^{\rho},\ \
q_{43\rho}=q_{23\rho}^{(1)}
\eeq

\beq
O_{42\rho}=x^2Z^{(2)}(x)Z^{(1)}_{\rho}(x,y),\ \
q_{42\rho}=q_{22\rho}^{(1)}
\eeq

\beq
O_{41\rho}^{(1)}=\frac{1}{4}x^2Z^{(12)}_{\rho}(x,y),\ \
q_{41\rho}^{(1)}=q_{21\rho}^{(1)}
\eeq
\beq
O_{41\rho}^{(2)}=\frac{\sqrt{3}}{4}xyZ^{(12)}_{\rho}(x,y),\ \
q_{41\rho}^{(2)}=q_{31\rho}^{(2)}
\eeq
\beq
O_{41\rho}^{(3)}=\frac{3}{4}y^2Z^{(12)}(x,y),\ \
q_{41\rho}^{(3)}=q_{21\rho}^{(3)}
\eeq

\beq
O_{53\rho}^{(1)}=\frac{1}{2}
x^2Z^{(2)}(x)U^{(2)}_{\rho}(x,y)(-1)^{\rho},\ \
q_{53\rho}^{(1)}=q_{23\rho}^{(1)}
\eeq
\beq
O_{53\rho}^{(2)}=\frac{\sqrt{3}}{2}
xyZ^{(2)}(x)U^{(2)}_{\rho}(x,y)(-1)^{\rho},\ \
q_{53\rho}^{(2)}=q_{23\rho}^{(2)}
\eeq

\beq
O_{52\rho}^{(1)}=(-1)^{\rho}O_{53\rho}^{(1)}
\ \
q_{52\rho}^{(1)}=q_{22\rho}^{(1)}
\eeq
\beq
O_{52\rho}^{(2)}=-(-1)^{\rho}O_{53\rho}^{(1)},\ \
q_{52\rho}^{(2)}=q_{22\rho}^{(2)}
\eeq

\beq
O_{51\rho}^{(1)}=-\frac{1}{4}x^2U^{(22)}_{\rho}(x,y),\ \
q_{51\rho}^{(1)}=q_{21\rho}^{(1)}
\eeq
\beq
O_{51\rho}^{(2)}=-\frac{\sqrt{3}}{4}xyU^{(22)}_{\rho}(x,y),\ \
q_{51\rho}^{(2)}=q_{21\rho}^{(2)}
\eeq
\beq
O_{51\rho}^{(3)}=\frac{3}{4}y^2U^{(22)}_{\rho}(x,y),\ \
q_{51\rho}^{(3)}=q_{21\rho}^{(3)}
\eeq

\beq
O_{63\rho}=O_{13\rho},\ \
q_{13\rho}=4\pi(\vsigma_3\cdot\vsigma_1\times\vsigma_2)(BD)_{\rho}
\eeq

\beq
O_{62\rho}=O_{12\rho}=,\ \
q_{12\rho}=4\pi(\vsigma_2\cdot\vsigma_3\vsigma_1)(BD)_{\rho}
\eeq

\beq
O_{61\rho}=O_{11\rho},\ \
q_{1\rho}=4\pi(\vsigma_1\cdot\vsigma_2\times\vsigma_3)(BD)_{\rho}
\eeq
%%%%%%%%%%%%%%%%%%%%%%%%%%%%%%%%%%%%%%%%%%%%%%%%%%%%%%%%%%%%%

\beq
O_{73\rho}=x^2Z^{(2)}{x}Z^{(1)}_\rho(x,y)(-1)^\rho,\ \
q_{73\rho}=
4\pi i\sqrt{2}\sum_{\mu\nu\epsilon\phi}C^{1\epsilon}_{1\mu 1\nu}
\sigma_3^{\epsilon}\sigma_1^{\mu}\sigma_2^{\phi}
 (B_{\nu\phi}D)_{\rho}
\eeq

\beq
O_{72\rho}^{(1)}=\frac{1}{4}x^2Z^{(1)}{x}Z^{(2)}_\rho(x,y),\ \
q_{72\rho}^{(1)}=
4\pi i\sqrt{2}\sum_{\mu\nu\epsilon\phi}C^{1\epsilon}_{1\mu 1\nu}
\sigma_2^{\epsilon}\sigma_3^{\mu}\sigma_1^{\phi}
 (B_{\nu\phi}D)_{\rho}
\eeq

\beq
O_{72\rho}^{(2)}=-\frac{\sqrt{3}}{4}xyZ^{(1)}(x)Z^{(2)}_\rho(x,y),\ \
q_{72\rho}^{(2)}=
4\pi i\sqrt{2}\sum_{\mu\nu\epsilon\phi}C^{1\epsilon}_{1\mu 1\nu}
\sigma_2^{\epsilon}\sigma_3^{\mu}\sigma_1^{\phi}
 (B_{\nu} D_\phi+B_{\phi}D_{\nu})_{\rho}
\eeq

\beq
O_{72\rho}^{(3)}=\frac{3}{4}y^2Z^{(1)}(x)Z^{(2)}_\rho(x,y),\ \
q_{72\rho}^{(3)}=
4\pi i\sqrt{2}\sum_{\mu\nu\epsilon\phi}C^{1\epsilon}_{1\mu 1\nu}
\sigma_2^{\epsilon}\sigma_3^{\mu}\sigma_1^{\phi}
 (BD_{\nu\phi})_{\rho}
\eeq
%%%%%%%%%%%%%%%%%%%%%%%%%%%%%%%%%%%%%%%%%%%%%%%

\beq
O_{71\rho}^{(1)}=\frac{1}{4}x^2Z^{(12)}_\rho(x,y),\ \
q_{71\rho}^{(1)}=
4\pi i\sqrt{2}\sum_{\mu\nu\epsilon\phi}C^{1\epsilon}_{1\mu 1\nu}
\sigma_1^{\epsilon}\sigma_2^{\mu}\sigma_3^{\phi}
 (B_{\nu\phi}D)_{\rho}
\eeq

\beq
O_{71\rho}^{(2)}=\frac{\sqrt{3}}{4}xyZ^{(12)}_\rho(x,y),\ \
q_{71\rho}^{(2)}=
4\pi i\sqrt{2}\sum_{\mu\nu\epsilon\phi}C^{1\epsilon}_{1\mu 1\nu}
\sigma_1^{\epsilon}\sigma_2^{\mu}\sigma_3^{\phi}
 (B_{\nu} D\phi+B_{\phi}D_{\nu})_{\rho}
\eeq

\beq
O_{71\rho}^{(3)}=\frac{3}{4}y^2Z^{(12)}_\rho(x,y),\ \
q_{72\rho}^{(3)}=
4\pi i\sqrt{2}\sum_{\mu\nu\epsilon\phi}C^{1\epsilon}_{1\mu 1\nu}
\sigma_1^{\epsilon}\sigma_2^{\mu}\sigma_3^{\phi}
 (BD_{\nu\phi})_{\rho}
\eeq
%%%%%%%%%%%%%%%%%%%%%%%%%%%%%%%%%%%%%%%%%%%%%%%%%%%%%%%%%%%%%%%
\beq
O_{83\rho}^{(1)}=-\frac{1}{4}x^2Z^{(1)}(x)Z^{(2)}_\rho(x,y)(-1)^{\rho},\ \
q_{83\rho}^{(1)}=
4\pi i\sqrt{2}\sum_{\mu\nu\epsilon\phi}C^{1\epsilon}_{1\mu 1\nu}
\sigma_3^{\epsilon}\sigma_2^{\mu}\sigma_1^{\phi}
 (B_{\nu\phi}D)_{\rho}
\eeq

\beq
O_{83\rho}^{(2)}=-\frac{\sqrt{3}}{4}xyZ^{(1)}{x}Z^{(2)}_\rho(x,y)(-1)^{\rho},\ \
q_{83\rho}^{(2)}=
4\pi i\sqrt{2}\sum_{\mu\nu\epsilon\phi}C^{1\epsilon}_{1\mu 1\nu}
\sigma_3^{\epsilon}\sigma_2^{\mu}\sigma_1^{\phi}
 (B_{\nu} D\phi+B_{\phi}D_{\nu})_{\rho}
\eeq

\beq
O_{83\rho}^{(3)}=-\frac{3}{4}y^2Z^{(1)}(x)Z^{(2)}_\rho(x,y)(-1)^{\rho},\ \
q_{83\rho}^{(3)}=
4\pi i\sqrt{2}\sum_{\mu\nu\epsilon\phi}C^{1\epsilon}_{1\mu 1\nu}
\sigma_3^{\epsilon}\sigma_2^{\mu}\sigma_1^{\phi}
 (BD_{\nu\phi})_{\rho}
\eeq
%%%%%%%%%%%%%%%%%%%%%%%%%%%%%%%%%%%%%%%%%%%%%%%%%%%%%%%%%%%%%%%%%%

\beq
O_{82\rho}=-x^2Z^{(2)}(x)Z^{(1)}_\rho(x,y),\ \
q_{82\rho}=
4\pi i\sqrt{2}\sum_{\mu\nu\epsilon\phi}C^{1\epsilon}_{1\mu 1\nu}
\sigma_2^{\epsilon}\sigma_3^{\mu}\sigma_1^{\phi}
 (B_{\nu\phi}D)_{\rho}
\eeq

%%%%%%%%%%%%%%%%%%%%%%%%%%%%%%%%%%%%%%%%%%%%%%%%%%%%%%%%%%%%%%%%%%
\beq
O_{81\rho}^{(1)}=-\frac{1}{4}x^2Z^{(21)}_\rho(x,y),\ \
q_{71\rho}^{(1)}=
4\pi i\sqrt{2}\sum_{\mu\nu\epsilon\phi}C^{1\epsilon}_{1\mu 1\nu}
\sigma_1^{\epsilon}\sigma_3^{\mu}\sigma_2^{\phi}
 (B_{\nu\phi}D)_{\rho}
\eeq

\beq
O_{81\rho}^{(2)}=\frac{\sqrt{3}}{4}xyZ^{(21)}_\rho(x,y),\ \
q_{81\rho}^{(2)}=
4\pi i\sqrt{2}\sum_{\mu\nu\epsilon\phi}C^{1\epsilon}_{1\mu 1\nu}
\sigma_1^{\epsilon}\sigma_3^{\mu}\sigma_2^{\phi}
 (B_{\nu} D\phi+B_{\phi}D_{\nu})_{\rho}
\eeq

\beq
O_{71\rho}^{(3)}=-\frac{3}{4}y^2Z^{(21)}_\rho(x,y),\ \
q_{72\rho}^{(3)}=
4\pi i\sqrt{2}\sum_{\mu\nu\epsilon\phi}C^{1\epsilon}_{1\mu 1\nu}
\sigma_1^{\epsilon}\sigma_3^{\mu}\sigma_2^{\phi}
 (BD_{\nu\phi})_{\rho}
\eeq

%%%%%%%%%%%%%%%%%%%%%%%%%%%%%%%%%%%%%%%%%%%%%%%%%%%%%%%%%%%%%%%%%
\beq
O_{93\rho}^{(1)}=
\frac{\sqrt{3}}{4}x^3Z^{(2)}(x)Z^{(2)}_{\rho}(x,y)(-1)^{\rho},\ \
q_{93\rho}^{(1)}
=4\pi i\sqrt{2}\sum_{\mu\nu\epsilon\phi\chi}C^{1\epsilon}_{1\mu 1\nu}
\sigma_3^{\epsilon}\sigma_1^{\phi}\sigma_2^{\chi}
(B_{\nu\phi\chi}D_{\mu})_{\rho}
\eeq

\beq
O_{93\rho}^{(2)}=
\frac{3}{4}x^2yZ^{(2)}(x)Z^{(2)}_{\rho}(x,y)(-1)^{\rho},\ \
q_{93\rho}^{(2)}
=4\pi i\sqrt{2}\sum_{\mu\nu\epsilon\phi\chi}C^{1\epsilon}_{1\mu 1\nu}
\sigma_3^{\epsilon}\sigma_1^{\phi}\sigma_2^{\chi}
(B_{\nu\chi}D_{\mu\phi})_{\rho}
\eeq
%%%%%%%%%%%%%%%%%%%%%%%%%%%%%%%%%%%%%%%%%%%%%%%%%%
\beq
O_{92\rho}^{(1)}=\frac{\sqrt{3}}{4}x^3Z^{(2)}(x)Z^{(2)}_{\rho}(x,y),\ \
q_{92\rho}^{(1)}
=4\pi i\sqrt{2}\sum_{\mu\nu\epsilon\phi\chi}C^{1\epsilon}_{1\mu 1\nu}
\sigma_2^{\epsilon}\sigma_3^{\phi}\sigma_1^{\chi}
(B_{\nu\phi\chi}D_{\mu})_{\rho}
\eeq

\beq
O_{92\rho}^{(2)}=\frac{3}{4}x^2yZ^{(2)}(x)Z^{(2)}_{\rho}(x,y),\ \
q_{93\rho}^{(2)}
=4\pi i\sqrt{2}\sum_{\mu\nu\epsilon\phi\chi}C^{1\epsilon}_{1\mu 1\nu}
\sigma_2^{\epsilon}\sigma_3^{\phi}\sigma_1^{\chi}
(B_{\nu\chi}D_{\mu\phi})_{\rho}
\eeq

%%%%%%%%%%%%%%%%%%%%%%%%%%%%%%%%%%%%%%%%%%%%%%%%%%%%%%%%%%%%%%
\beq
O_{91\rho}^{(1)}=-\frac{\sqrt{3}}{8}x^3yZ^{(22)}_{\rho}(x,y),\ \
q_{91\rho}^{(1)}
=4\pi i\sqrt{2}\sum_{\mu\nu\epsilon\phi\chi}C^{1\epsilon}_{1\mu 1\nu}
\sigma_1^{\epsilon}\sigma_2^{\phi}\sigma_3^{\chi}
(B_{\nu\phi\chi}D_{\mu})_{\rho}
\eeq

\beq
O_{91\rho}^{(2)}=\frac{3}{8}x^2y^2Z^{(22)}_{\rho}(x,y)),\ \
q_{91\rho}^{(2)}
=4\pi i\sqrt{2}\sum_{\mu\nu\epsilon\phi\chi}C^{1\epsilon}_{1\mu 1\nu}
\sigma_1^{\epsilon}\sigma_2^{\phi}\sigma_3^{\chi}
(B_{\nu\chi}D_{\mu\phi}-B_{nu\phi}D_{\mu\chi})_{\rho}
\eeq

\beq
O_{91\rho}^{(3)}=\frac{3\sqrt{3}}{8}xy^3Z^{(22)}_{\rho}(x,y),\ \
q_{91\rho}^{(3)}
=4\pi i\sqrt{2}\sum_{\mu\nu\epsilon\phi\chi}C^{1\epsilon}_{1\mu 1\nu}
\sigma_1^{\epsilon}\sigma_2^{\phi}\sigma_3^{\chi}
(B_{\nu}D_{\mu\phi\chi})_{\rho}
\eeq
%%%%%%%%%%%%%%%%%%%%%%%%%%%%%%%%%%%%%%%%%%%%%%%%%%%%%%%%%%%%%%%%%%

\section{Geometrical factors in the MGL basis}
To pass to the MGL basis we have only to transform matrix elements of
operators $q$ in the $a$-basis with the help of coefficient functions $F$:
\beq
<\alpha'|q_{ik\rho}^{(l)}|\alpha>=
\sum_{a',a}F^{a'}_{\alpha'}<a'|q_{ik\rho}^{(l)}|a>F^{a}_{\alpha}
\label{melal}
\eeq
This expression involves many summations over projection of momenta.
Possibly they can be done analytically, using the appropriate
formulas from Varshalovich.  To this aim one should use representation
of matrices Pauli in terms of Clebsh-Gordon coefficients:
\beq
\sigma^{\mu}_{p'p}=\sqrt{2}(-1)^{1/2+\mu-p}C^{1\mu}_
{\frac{1}{2}-p'\frac{1}{2}p},\ \
[\sigma_{\mu}]_{p'p}=\sqrt{2}(-1)^{1/2-p}C^{1\mu}_
{\frac{1}{2}p'\frac{1}{2}-p}
\eeq
so that the whole expression (\ref{melal}) turns out to be a product
of Clebsh-Gordon coefficients summed over momentum projections.
Since there are many operators $q_{ik\rho}^{(l)}$ to be processed
in this way, probably it is more practical to do the sums
numerically on the computer to calculate matrix elements (\ref{melal})
as a preliminary for dynamical calculations.

In the following, just for illustration, we shall try to do the sum over projections
analytically for some simplest operators $q$.

%\newpage
\subsection{Operators $O_1$}
We are going to study the matrix element
\beq
<\alpha'|q_{13\rho}|\alpha>=4\pi<\alpha'|(\vsigma_1\vsigma_2)(BD)_{\rho}|\alpha>
\label{alele}
\eeq
First let us write out the relevant factors
\beq
F^{a'}_{\alpha'}F^a_\alpha=
C^{MM_z}_{s's'_z\lambda'\lambda'_z}C^{s's'_z}_{\frac{1}{2}p'_1J'J'_z}C^{J'J'_z}_{\sigma'\sigma'_zl'l'_z}
C^{\sigma'\sigma'_z}_{\frac{1}{2}p'_2\frac{1}{2}p'_3}
C^{MM_z}_{ss_z\lambda\lambda_z}C^{ss_z}_{\frac{1}{2}p_1JJ_z}C^{JJ_z}_{\sigma\sigma_zll_z}
C^{\sigma\sigma_z}_{\frac{1}{2}p_2\frac{1}{2}p_3}
\label{ffac}
\eeq
\[
<a'|q_{13\rho}|a>=2(-1)^{1-p_1-p_2}
\frac{\prod_{l'\rho}}{\prod_l}C^{l0}_{l'0\rho 0}
\frac{\prod_{\lambda\rho}}{\prod_\lambda'}C^{\lambda'l0}_{\lambda0\rho 0}\delta_{p'_3p_3}
\]\beq
\sum_{\mu\tau}(-1)^{\mu}
C^{1\mu}_{\frac{1}{2}-p'_1\frac{1}{2}p_1}C^{1\mu}_{\frac{1}{2}p'_2\frac{1}{2}-p_2}
C^{ll_z}_{l'l'_z\rho\tau}
C^{\lambda'\lambda'_z}_{\lambda\lambda_z\rho\tau}
\label{aele}
\eeq
Multiplying (\ref{ffac}) and (\ref{aele}) we obtain an explicit expression
for the matrix element (\ref{alele})
\[
<\alpha'|q_{13\rho}|\alpha>=2
\frac{\prod_{l'\rho}}{\prod_l}C^{l0}_{l'0\rho 0}
\frac{\prod_{\lambda\rho}}{\prod_\lambda'}C^{\lambda'l0}_{\lambda0\rho 0}\]\[
\sum_{\lambda'_z,l'_z,p'_1,p'_2}\sum_{\lambda_z,l_z,p_1,p_2}\sum_{\mu,\tau,p_3}
(-1)^{1-p_1-p_2+\mu}\]\beq
C^{MM_z}_{s's'_z\lambda'\lambda'_z}C^{s's'_z}_{\frac{1}{2}p'_1J'J'_z}C^{J'J'_z}_{\sigma'\sigma'_zl'l'_z}
C^{\sigma'\sigma'_z}_{\frac{1}{2}p'_2\frac{1}{2}p_3}
C^{MM_z}_{ss_z\lambda\lambda_z}C^{ss_z}_{\frac{1}{2}p_1JJ_z}C^{JJ_z}_{\sigma\sigma_zll_z}
C^{\sigma\sigma_z}_{\frac{1}{2}p_2\frac{1}{2}p_3}
C^{1\mu}_{\frac{1}{2}-p'_1\frac{1}{2}p_1}C^{1\mu}_{\frac{1}{2}p'_2\frac{1}{2}-p_2}
C^{ll_z}_{l'l'_z\rho\tau}
C^{\lambda'\lambda'_z}_{\lambda\lambda_z\rho\tau}
\label{alelexpl}
\eeq

We can perform partial summation over $p_2,p'_2, p_3$
\[
S_1=\sum_{p_2,p'_2, p_3}(-1)^{1/2-p_2}C^{\sigma\sigma_z}_{\frac{1}{2}p_2\frac{1}{2}p_3}
C^{\sigma'\sigma'_z}_{\frac{1}{2}p'_2\frac{1}{2}p_3}
C^{1\mu}_{\frac{1}{2}p'_2\frac{1}{2}-p_2}\]\beq=
(-1)^{\sigma+\sigma'}
\sum_{p_2,p'_2, p_3}(-1)^{1/2-p_2}C^{\sigma\sigma_z}_{\frac{1}{2}p_3\frac{1}{2}p_2}
C^{\sigma'\sigma'_z}_{\frac{1}{2}p_3\frac{1}{2}p'_2}
C^{1\mu}_{\frac{1}{2}p'_2\frac{1}{2}-p_2}
\eeq
We use Varshalovich 8.7.3.(17) with
\[b\beta=\frac{1}{2}p_3,\ \ a\alpha=\frac{1}{2}p_2,\ \ c\gamma=\sigma\sigma_z,\ \
d\delta=\frac{1}{2}p'_2,\ \ e\epsilon=\sigma'\sigma'_z,\ \ f\phi=1,\mu\]
to obtain
\beq
S_1=(-1)^{-\sigma}\prod_{\sigma 1}C^{\sigma'\sigma'_z}_{\sigma\sigma_z 1\mu}
  \left \{ \begin{array}{ccc}
      \frac{1}{2}   & \frac{1}{2} & \sigma \\
    \sigma' & 1  & \frac{1}{2}
    \end{array} \right \}
\eeq

As a next step we may try to sum
\beq
S_2= (-1)^{-\mu}\sum_{l_z,l'_z,\sigma_z,\sigma'_z}
C^{J'J'_z}_{\sigma'\sigma'_zl'l'_z} C^{JJ_z}_{\sigma\sigma_zll_z}
C^{ll_z}_{l'l'_z\rho\tau} C^{\sigma'\sigma'_z}_{\sigma\sigma_z 1\mu}
\label{sum2}
\eeq
We note that
\[\mu=\sigma'_z=\sigma_z,\ \ \sigma_z=J_z-l_z,\ \ {\rm so\ \ that}\ \
\mu=\sigma'_z+l_z-J_z\] and (\ref{sum2}) can be rewritten as
\beq
S_2= (-1)^{-J_z-l-\sigma'}
\sum_{l_z,l'_z,\sigma_z,\sigma'_z}(-1)^{l-l_z+\sigma'-\sigma'_z}
C^{J'J'_z}_{\sigma'\sigma'_zl'l'_z} C^{JJ_z}_{\sigma\sigma_zll_z}
C^{ll_z}_{l'l'_z\rho\tau} C^{\sigma\sigma'_z}_{\sigma\sigma_z 1\mu}
\label{sum21}
\eeq
This sum corresponds to Varshalovich 8.7.4(30) with
\[a\alpha=\rho\tau,\ \ b\beta=l'l'_z,\ \ c\gamma=ll_z,\ \
d\delta=1\mu,\ \ e\epsilon=\sigma'\sigma'_z,\ \ f\phi=\sigma\sigma_z,\ \
g\eta=J'J'_z,\ \ j\mu=JJ_z
\]
Taking into account factors arising due to replacements in lower arguments
in Clebsh-Gordon coefficients we find
\beq
S_2=(-1)^{l+l'-J-\mu-\tau-J_z}\prod_{l\sigma'JJ'}
\sum_{k\kappa}C^{k\kappa}_{J'J'_zJ-J_z}C^{k\kappa}_{1\mu\rho -\tau}
  \left \{ \begin{array}{ccc}
    l   & l' & \rho \\
    \sigma & \sigma'  & 1\\
    J  & J' & k
    \end{array} \right \}
\eeq

Now we sum
\beq
S_3=(-1)^{1/2-p_1-J_z}\sum_{J_zJ'_zp_1p'_1}
C^{s's'_z}_{\frac{1}{2}p'_1J'J'_z}C^{ss_z}_{\frac{1}{2}p_1JJ_z}C^{k\kappa}_{J'J'_zJ-J_z}
C^{1\mu}_{\frac{1}{2}-p'_1\frac{1}{2}p_1}
\eeq
We have $p_1+J_z=s_z$ and
\[C^{1\mu}_{\frac{1}{2}-p'_1\frac{1}{2}p_1}=C^{1-\mu}_{\frac{1}{2}p'_1\frac{1}{2}-p_1}\]
so that
\beq
S_3=(-1)^{1/2-s_z}\sum_{J_zJ'_zp_1p'_1}
C^{s's'_z}_{\frac{1}{2}p'_1J'J'_z}C^{ss_z}_{\frac{1}{2}p_1JJ_z}C^{k\kappa}_{J'J'_zJ-J_z}
C^{1-\mu}_{\frac{1}{2}p'_1\frac{1}{2}-p_1}
\eeq
We use Varshalovich 8.7.4.(21) with
\[a\alpha=k\kappa,\ \ b\beta=J'J'_z,\ \ c\gamma=JJ_z,\ \ d\delta=1-\mu,\ \
e\epsilon=\frac{1}{2}p'_1,\ \ f\phi=\frac{1}{2}p_1,\ \ g\eta=s's'_z,\ \
j\mu=ss_z\]
Taking into account factors due to interchanges of lower indexes we find
\beq
S_3=(-1)^{J-2s'-s-s_z}\prod_{k1s's}\ \sum_{k_1\kappa_1}
C^{k_1\kappa_1}_{s's'_zs-s_z}C^{k_1\kappa_1}_{1-\mu k\kappa}
  \left \{ \begin{array}{ccc}
    J   & J' & k \\
    \frac{1}{2} & \frac{1}{2}  & 1\\
    s  & s' & k_1
    \end{array} \right \}
\eeq

Collecting our results, at this step we have
\[
<\alpha'|q_{13\rho}|\alpha>=6
\prod_{l'\rho\rho\lambda\sigma\sigma' JJ'kss'}{\prod}^{-1}_{\lambda'}
C^{l0}_{l'0\rho 0}
C^{\lambda'l0}_{\lambda0\rho 0}\]\[
 \left \{ \begin{array}{ccc}
      \frac{1}{2}   & \frac{1}{2} & \sigma \\
    \sigma' & 1  & \frac{1}{2}
    \end{array} \right \}
  \left \{ \begin{array}{ccc}
    l   & l' & \rho \\
    \sigma & \sigma'  & 1\\
    J  & J' & k
    \end{array} \right \}
  \left \{ \begin{array}{ccc}
    J   & J' & k \\
    \frac{1}{2} & \frac{1}{2}  & 1\\
    s  & s' & k_1
    \end{array} \right \}
 \]\beq
\sum_{\lambda'_z,s'_z,\lambda_z,s_z,\mu,\tau,\kappa\kappa_1}
(-1)^{-\sigma+l+l'-2s'-s-\mu-\tau-s_z}
C^{MM_z}_{s's'_z\lambda'\lambda'_z}
C^{MM_z}_{ss_z\lambda\lambda_z}C^{k\kappa}_{1\mu\rho -\tau}
C^{\lambda'\lambda'_z}_{\lambda\lambda_z\rho\tau}
C^{k_1\kappa_1}_{s's'_zs-s_z}C^{k_1\kappa_1}_{1-\mu k\kappa}
\label{alelexp2}
\eeq

Now we sum
\beq
S_4=\frac{1}{2M+1}\sum_{M_z,\lambda_z,\lambda'_z}
C^{MM_z}_{s's'_z\lambda'\lambda'_z}
C^{MM_z}_{ss_z\lambda\lambda_z}C^{\lambda'\lambda'_z}_{\lambda\lambda_z\rho\tau}
\eeq
We transform
\beq
C^{\lambda'\lambda'_z}_{\lambda\lambda_z\rho\tau}=(-1)^{\lambda-\lambda_z}
\prod_{\lambda'}{\prod}^{-1}_{\rho}C^{\rho-\tau}_{\lambda\lambda_z\lambda'\lambda'-\lambda'_z}
\eeq
to transform
\beq
S_4=(-1)^{\lambda+\lambda'-\rho}\prod_{\lambda'}{\prod}^{-1}_{\rho}
\frac{1}{2M+1}\sum_{M_z,\lambda_z,\lambda'_z}(-1)^{\lambda-\lambda_z}
C^{MM_z}_{s's'_z\lambda'\lambda'_z}
C^{MM_z}_{ss_z\lambda\lambda_z}C^{\rho\tau}_{\lambda-\lambda_z\lambda'\lambda'_z}
\eeq
We use Varshalovich 8.7.3.16 with
\[a\alpha=\lambda'\lambda'_z,\ \ b\beta=\lambda-\lambda_z,\ \ c\gamma=\rho\tau,\ \
d\delta=MM_z,\ \ e\epsilon=ss_z,\ \ f\phi=s's'_z\]
to obtain
\beq
S_4=(-1)^{\lambda'+M+s'}\prod_{\lambda'}{\prod}^{-1}_{s}C^{ss_z}_{\rho\tau s's'_z}
 \left \{ \begin{array}{ccc}
      \lambda'   & \lambda & \rho \\
    s & s'  & M
    \end{array} \right \}
\eeq
Our next sum is
\beq
S_5=\sum_{\tau,\kappa,\kappa_1,s_z}
(-1)^{-\tau-s_z}C^{k_1\kappa_1}_{s's'_zs-s_z}C^{k_1\kappa_1}_{1-\mu k\kappa}
C^{ss_z}_{\rho\tau s's'_z}C^{k\kappa}_{1\mu\rho -\tau}
\eeq
We have $\tau+s'_z=s_z$ so that $(-1)^{-\tau-s_z}=(-1)^{\tau-s_z}=(-1)^{-s'_z}$
and can be taken out of the sum.
The rest corresponds to Varshalovich 8.7.4.(24) with
\[a\alpha=1\mu,\ \ b\beta=\rho\tau,\ \ c\gamma=k-\kappa,\ \
d\delta=s'-s'_z,\ \ e\epsilon=s-s_z,\ \ f\phi=k_1\kappa_1,\ \ g\eta=s's'_z,
\ \ ,j\mu=1-\mu\]
We obtain
\beq
S_5=(-1)^{-s'_z}(-1)^{\rho-k-s'-s'_z-\mu}\prod_{ksk_1k_1}
\sum_{k_2\kappa_2}C^{k_2\kappa_2}_{s'-s'_z1-\mu}C^{k_2\kappa_2}_{s'-s'_z1-\mu}
  \left \{ \begin{array}{ccc}
    k  & \rho & 1 \\
    k_1 & s  & s'\\
    1  & s' & k_2
    \end{array} \right \}
\eeq
The last sum over $s'_z$ and $\mu$ cannot be done because of the factor $(-1)^{-2s'_z}$
( $(-1)^{-\mu}$ is cancelled by the same factor in (\ref{alelexp2})),
which is strange.

To check we shall try to use an alternative formula Varshalovich 8.7.4.(20) transforming the
relevant Clebsh-Gordon coefficients to the form corresponding to this formula
(putting all the projections to be summed down).
We have
\beq S_5=(-1)^{-s'_z}\tilde{S}_5
\label{stilde}
 \eeq
where
\beq
\tilde{S}_5=\sum_{\tau,\kappa,\kappa_1,s_z}
C^{k_1\kappa_1}_{s's'_zs-s_z}C^{k_1\kappa_1}_{1-\mu k\kappa}
C^{ss_z}_{\rho\tau s's'_z}C^{k\kappa}_{1\mu\rho -\tau}
\eeq
Now
\[
C^{k_1\kappa_1}_{s's'_zs-s_z}=(-1)^{s'+s-k+s+s_z}\sqrt{\frac{2k_1+1}{2s'+1}}
C^{s'-s'_z}_{s-s_z k_1-\kappa_1}\]
\[
C^{k_1\kappa_1}_{1-\mu k\kappa}=(-1)^{1+k+k_1+k-\kappa}\sqrt{\frac{2k_1+1}{3}}
C^{1\mu}_{k\kappa k_1-\kappa_1}
\]
\[
C^{ss_z}{\rho\tau s's'_z}=(-1)^{\rho-\tau}\sqrt{\frac{2s+1}{2s'+1}}
C^{s'-s'_z}_{\rho\tau s-s_z}
\]
\[
C^{k\kappa}_{1\mu\rho -\tau}=(-1)^{1+\rho-k+\rho+\tau}\sqrt{\frac{2k+1}{2\rho +1}}
C^{1-\mu}_{\rho -\tau k-\kappa}=(-1)^{\rho+k-1}C^{1\mu}_{\rho \tau k\kappa}
\]
So we get
\beq
\tilde{S}_5=(-1)^{s'+2s-k_1+\rho+k-1}\prod_{k_1k_1sk}{\prod}^{-1}_{s'1s'\rho}\Big)^{-1}
\sum_{\tau,\kappa,\kappa_1,s_z}(-1)^{-s_z-\kappa}C^{s'-s'_z}_{s-s_z k_1-\kappa_1}
C^{1\mu}_{k\kappa k_1-\kappa_1}C^{s'-s'_z}_{\rho\tau s-s_z}
C^{1\mu}_{\rho \tau k\kappa}
\eeq
However we have
\[\kappa_1=s'_z-s_z=\kappa-\mu,\ \ {\rm so\ \ that}\ \ \kappa=s'_z-s_z+\mu\]
and also obviously $\kappa$ is an integer and
\[(-1)^{-s_z-\kappa}=(-1)^{-s_z+\kappa}=(-1)^{s'_z+\mu}\].
So we can take the sign factor out of the sum, where $(-1)^{s_z}$ cancels with the sign factor
in (\ref{stilde}).
The sum itself corresponds to Varshalovich. 8.7.4.(20) with
\[ a\alpha=s'-s'_z ,\ \ b\beta=ss_z,\ \ c\gamma=k_1\kappa_1,\ \ d\delta=1\mu,\ \
e\epsilon=\rho\tau,\ \ f\phi=k\kappa,\ \ g\eta=s'-s'_z,\ \ j\mu=1\mu\]
Summation gives
\beq
\tilde{S}_5=(-1)^{s'+2s-k_1+k-1}(-1)^{s_z+\mu}\prod_{k_1k_1sk1}
\sum_{k_2\kappa_2}C^{k_2\kappa_2}_{s'-s'_z1\mu}C^{k_2\kappa_2}_{1\mu s'-s'_z}
  \left \{ \begin{array}{ccc}
    k_1  & s & s'\\
    k & \rho  & 1\\
    1  & s' & k_2
    \end{array} \right \}
\eeq
Factor $(-1)^{-\mu}$ cancels with the same factor in (\ref{alelexp2}).
and $(-1)^{s'_z}$ with the inverse factor in (\ref{stilde}), so that no
sign factors depending on projections $s_z$ and $\mu$ remain.
This allows to do the last summation over $s_z$ and $\mu$:
\beq
S_6=\sum_{s_z,\mu}C^{k_2\kappa_2}_{s'-s'_z1\mu}C^{k_2\kappa_2}_{1\mu s'-s'_z}=1
\eeq

Collecting all the factors, we get
\[
<\alpha'|q_{13\rho}|\alpha>=-6\sqrt{3}
\prod_{l'\rho\rho\lambda\sigma\sigma' JJ'ss'}
C^{l0}_{l'0\rho 0}(-1)^{l+l'-\sigma+\lambda'+\rho+M+s'+2s}
C^{\lambda'l0}_{\lambda0\rho 0}\]\[\sum_{k,k_1,k_2}(-1)^{k-k_1}
\prod_{kkk_1k_1}\]\beq
 \left \{ \begin{array}{ccc}
      \frac{1}{2}   & \frac{1}{2} & \sigma \\
    \sigma' & 1  & \frac{1}{2}
    \end{array} \right \}
  \left \{ \begin{array}{ccc}
    l   & l' & \rho \\
    \sigma & \sigma'  & 1\\
    J  & J' & k
    \end{array} \right \}
  \left \{ \begin{array}{ccc}
    J   & J' & k \\
    \frac{1}{2} & \frac{1}{2}  & 1\\
    s  & s' & k_1
    \end{array} \right \}
    \left \{ \begin{array}{ccc}
      \lambda'   & \lambda & \rho \\
    s & s'  & M
    \end{array} \right \}
      \left \{ \begin{array}{ccc}
    k_1  & s & s'\\
    k & \rho  & 1\\
    1  & s' & k_2
    \end{array} \right \}
\label{alelexp3}
\eeq
This expression should serve as a test for the numerical code to
do the summation over projections numerically.
%%%%%%%%%%%%%%%%%%%%%%%%%%%%%%%%%%%%%%%%%%%%%%%%%%%%%%%%%%%%%%%%%
\section*{Acknowledgments}
This work is supported by the NSF (HRD-0833184) and NASA (NNX09AV07A)

\end{document}